\documentclass[%
reprint,
superscriptaddress,
showpacs,
preprintnumbers,
amsmath,
amssymb,
aps,
prd,
]{revtex4-1}

\usepackage{graphicx}
\usepackage{dcolumn}
\usepackage{bm}

\begin{document}

\title{Radially excited $U\left(1\right)$ gauged $Q$-balls}
\author{A. Yu. Loginov}
\email{aloginov@tpu.ru}
\affiliation{Tomsk State University of Control Systems and Radioelectronics, 634050 Tomsk, Russia}
\author{V. V. Gauzshtein}
\affiliation{Tomsk Polytechnic University, 634050 Tomsk, Russia}

\date{\today}

\begin{abstract}
Radially excited  $U(1)$ gauged  $Q$-balls  are  studied  using both analytical
and numerical methods.
Unlike the nongauged case,  there  exists  only  a  finite  number  of radially
excited gauged  $Q$-balls at given values of the model's parameters.
Similarly to the unexcited   gauged  $Q$-ball, the  radially excited one cannot
possess the Noether charge exceeding some limiting value.
This  limiting  Noether  charge  decreases  with  an  increase  in  the  radial
excitation of the gauged $Q$-ball.
For $n$-th radial excitation, there is a  maximum  allowable value of the gauge
coupling constant,  and  the  existence  of  the $n$-th radially excited gauged
$Q$-ball  becomes  impossible  if  the  gauge  coupling  constant  exceeds this
limiting value.
Similarly to the limiting Noether charge, the limiting gauge  coupling constant
decreases with an increase in the radial excitation.
At a fixed Noether charge, the  energy  of  the  gauged $Q$-ball increases with
an increase in the radial  excitation,  and  thus the radially  excited  gauged
$Q$-ball is unstable  against  transit  into a less  excited  or unexcited one.
\end{abstract}

\maketitle

\section{\label{sec:I} Introduction}

In  the  framework  of   field   theory,  solitons   are  spatially  localized,
nonspreading  solutions   to   field   equations   possessing   finite  energy.
Solitons play  an  important  role  in  high  energy  physics, condensed matter
physics, cosmology, and  hydrodynamics.
They  can  be   divided   into   two  main  groups:  topological  solitons  and
nontopological solitons.
The existence and stability of topological solitons result from the topological
nontriviality of their field configurations \cite{Manton}.
This nontriviality  means  that  a  topological  soliton cannot be smoothly and
continuously   deformed   into   the   vacuum   field  configuration,  which is
topologically trivial.
Hence,  there  exists  an  infinite   potential barrier between the topological
soliton  and  the  vacuum  state,  and  the  soliton  cannot decay into a field
configuration in  the  functional  neighbourhood  of  the topologically trivial
vacuum.

In  contrast,  the  field   configurations  of   nontopological   solitons  are
topologically  trivial,  and   the  existence  of  nontopological  solitons  is
therefore due to the dynamics of corresponding field models \cite{lee}.
In particular, nontopological solitons   exist  in  field  models that  possess
global  symmetries   (which    may    be    both   Abelian   and   non-Abelian)
\cite{rosen, fried, coleman, saf1, saf2} and  have interaction potentials  that
meet certain conditions \cite{lee, paccetti}.
The basic  property  of  a  nontopological  soliton  is that it is the extremum
(minimum or saddle point) of  the  energy  functional  at  a fixed value of the
Noether  charge  corresponding  to  the  global symmetry group generator of the
field model.
This feature  of  nontopological  solitons  results in  the characteristic time
dependence $\propto \exp \left( - i\omega t\right)$  of their fields.
The time dependence of the  soliton's field  allows  the severe restrictions of
Derrick's  theorem  \cite{derrick} to be avoided,  meaning that  nontopological
solitons composed of scalar fields can  exist  in  space-time with an arbitrary
number of  spatial dimensions.

The simplest of  nontopological  solitons  is  the $Q$-ball  \cite{coleman}, a
coherent state of a self-interacting complex scalar field.
$Q$-balls exist in models of complex scalar fields possessing $U\left(1\right)$
global symmetry and  certain self-interaction potentials.
The characteristic   feature  of  nontopological  solitons  in  general  and of
$Q$-balls in  particular  is  the  presence  of  an infinite number of radially
excited  states  in  addition  to  the    basic   unexcited  state \cite{fried,
volkov_2002, kunz_2005, mai_2012}.
The basic and radially  excited  states of $Q$-balls are spherically symmetric.
The profile function of the basic unexcited  state  of  a $Q$-ball has no nodes
for  any  finite  radius,  whereas  the  $n$-th  radially  excited  state  of a
$Q$-ball has exactly $n$ nodes at finite radii.

When the  $U(1)$  symmetry  of  the model  is global, the Noether charge of the
$Q$-ball corresponds to the particle number.
However, the global $U(1)$ symmetry can be gauged by means of the Abelian gauge
field,  which  interacts  minimally  with  the  model's  complex  scalar field.
As  in  the  case of  global  $U(1)$  symmetry,  the $U(1)$  gauged   models of
complex  self-interacting  scalar  fields  admit  the  existence  of  $Q$-balls
\cite{rosen2, klee, lee_yoon_1991, anag, levi, ardoz_2009,  benci, tamaki_2014,
gulamov_2014, brihaye_prd_2014, hong_prd_2015, gulamov_2015}.
These gauged $Q$-balls are  electrically  charged  objects,  and thus possess a
long-range electric field.
The electric charge of gauged $Q$-balls  is  the product of the  Noether charge
and the gauge coupling constant, which defines the interaction strength between
the gauge field and the complex scalar field.
The properties of gauged $Q$-balls differ significantly from those of nongauged
ones.
In particular, the Noether (electric) charge and the energy of gauged $Q$-balls
cannot be arbitrarily large \cite{klee, gulamov_2015} provided that the  second
derivative  of  the self-interaction  potential $d^{2} V\left( \left\vert  \phi
\right\vert \right)/d\left\vert   \phi   \right\vert^{2}$   is   finite  at the
origin \cite{tamaki_2014}. 
A gauged $Q$-ball also cannot exist if the gauge coupling constant exceeds some
maximum value \cite{klee}, which  depends on the other parameters of the model.

As in the  nongauged  case,  there  exist  radially  excited  gauged $Q$-balls.
These also cannot possess an arbitrarily large
electric charge, and can  exist  only  if  the gauge coupling constant does not
exceed some limiting value; both the maximum  possible  Noether  charge and the
maximum possible  gauge  coupling  constant  decrease  with  an increase in the
radial excitation of a gauged $Q$-ball.
Furthermore, there is only a finite number of radially excited gauged $Q$-balls
for given  values  of  the  gauge  coupling  constant  and the other parameters
of the model.
In this  paper, we  study  radially  excited  gauged  $Q$-balls  from  both  an
analytical and numerical perspective.
We ascertain the basic properties of these objects  and discuss related issues.

The  paper  is  structured as follows.
In Sec.~\ref{sec:II},  we  describe  briefly  the  Lagrangian,  the symmetries,
and the field equations of the model under consideration.
In  Sec.~\ref{sec:III}, the general properties  of   the  gauged  $Q$-balls are
considered and discussed.
In Sec.~\ref{sec:IV},  we present numerical results for radially excited gauged
$Q$-balls.
Some numerical results relating to unexcited gauged $Q$-balls are also included
in Sec.~\ref{sec:IV} for completeness.
In the final section, we  briefly  summarize the results obtained in this work.
Three appendices are also included.
In Appendix A, we show that the  gauged $Q$-ball's fields do not depend on time
in the unitary gauge, and  establish  the basic relation between the energy and
the  Noether  charge of the gauged $Q$-ball.
In Appendix B, we discuss the reason for  the existence of the maximum possible
electric charge for gauged $Q$-balls.
In Appendix C,  we  ascertain  the  reason  for the existence of the inflection
point in the  curve  describing  the  dependence  of  the  energy of the gauged
$Q$-ball on the Noether charge.

Throughout this paper, we use the natural units $c = 1$, $\hbar = 1$.

\section{\label{sec:II}   Lagrangian and field equations of the model}

The  gauge   model   we   are   interested   in   has  the  Lagrangian  density
\begin{equation}
\mathcal{L}=-\frac{1}{4}F_{\mu \nu }F^{\mu \nu }+\left( D_{\mu }\phi \right)
^{\ast }D^{\mu }\phi -V\left( \left\vert \phi \right\vert \right). \label{II:1}
\end{equation}
This model describes the self-interacting complex scalar field $\phi$ minimally
interacting with  the  Abelian  gauge  field  $A_{\mu}$  through  the covariant
derivative
\begin{equation}
D_{\mu }\phi = \partial_{\mu }\phi + i e A_{\mu }\phi.             \label{II:2}
\end{equation}
The local gauge transformations
\begin{eqnarray}
\phi \left( x\right)  &\rightarrow &\phi ^{\prime }\left( x\right) =
\exp\left( -ie\Lambda \left( x\right) \right) \phi \left( x\right) ,  \nonumber
\\
A_{\mu }\left( x\right)  &\rightarrow &A_{\mu }^{\prime }\left( x\right)
=A_{\mu }\left( x\right) +\partial _{\mu }\Lambda \left(x\right)  \label{II:3}
\end{eqnarray}
leave invariant the Lagrangian density (\ref{II:1}).
A special case of Eq.~(\ref{II:3}) is the  global  phase  transformations $\phi
\left( x\right) \rightarrow \phi ^{\prime }\left( x\right) =\exp \left(-i\alpha
\right) \phi \left( x\right)$.
The invariance of the Lagrangian  density (\ref{II:1}) under these global phase
transformations leads to the conserved Noether current of the model
\begin{equation}
j_{N}^{\nu } = i \left[ \phi ^{\ast }D^{\nu }\phi - \left( D^{\nu} \phi \right)
^{\ast }\phi \right].                                              \label{II:4}
\end{equation}

The self-interaction of the complex scalar field  is described by the six-order
potential
\begin{equation}
V\left( \left\vert \phi \right\vert \right) = m^{2}\left\vert \phi
\right\vert ^{2} - \frac{g}{2}\left\vert \phi \right\vert ^{4}+\frac{h}{3}
\left\vert \phi \right\vert^{6},                                   \label{II:5}
\end{equation}
where the self-interaction coupling  constants  $g$  and  $h$ are assumed to be
positive.
We assume that the  potential $V\left( \left\vert \phi \right\vert \right)$ has
a global minimum at $\phi = 0$  and thus there is no spontaneously broken gauge
symmetry.
For this to hold,  the  parameters  of  the  potential in Eq.~(\ref{II:5}) must
satisfy the inequality $3g^{2} < 16hm^{2}$.

By varying the action $S=\int \mathcal{L}d^{3}xdt$ in the corresponding fields,
we obtain the field equations of the model:
\begin{align}
D_{\mu }D^{\mu }\phi +m^{2}\phi -g\left\vert \phi \right\vert ^{2}\phi
+h\left\vert \phi \right\vert ^{4}\phi & = 0,                     \label{II:6a}
\\
\partial _{\mu }F^{\mu \nu }& = j^{\nu},                          \label{II:6b}
\end{align}
where the electromagnetic current density $j^{\nu} = e j_{N}^{\nu }$.
Later on, we shall also need the expression for  the energy-momentum tensor for
a field configuration of the model:
\begin{eqnarray}
T_{\mu \nu} &=&-F_{\mu \lambda}F_{\nu }^{\;\lambda}+\frac{1}{4}\eta_{\mu \nu}
F_{\lambda \rho }F^{\lambda \rho }                             \nonumber
\\
&&+\left( D_{\mu }\phi \right) ^{\ast }D_{\nu }\phi +\left( D_{\nu }\phi
\right) ^{\ast }D_{\mu }\phi                                   \nonumber
\\
&&-\eta _{\mu \nu }\left( \left( D_{\mu }\phi \right) ^{\ast }D^{\mu }\phi
-V\left( \left\vert \phi \right\vert \right) \right),              \label{II:7}
\end{eqnarray}
where the metric tensor $\eta_{\mu \nu} = \text{diag}\left(+1,-1,-1,-1\right)$.

\section{\label{sec:III} Some properties of gauged $Q$-balls}

By definition, the  $Q$-ball field configuration is an extremum  of  the energy
functional $E=\int T_{00}d^{3}x$ at  a fixed value of the Noether charge $Q_{N}
 = \int j_{N}^{0}d^{3}x$.
Thus, the $Q$-ball field configuration is  a conditional extremum of the energy
functional.
According to Lagrange's method of multipliers, the $Q$-ball is an unconditional
extremum of the  functional $F  =  E  -  \lambda Q_{N}$, where $\lambda$ is the
Lagrange multiplier.
Using Lagrange's  method  of  multipliers  and the Hamilton field equations, we
show in  Appendix A that the  $Q$-ball field  configuration  does not depend on
time in the unitary gauge $\text{Im}\left( \phi \right) = 0$.
Next, we assume that in the unitary  gauge, the $Q$-ball field configuration is
spherically symmetric.
The spherical symmetry,  time  independence,  and  regularity  of  the $Q$-ball
field configuration lead to  the  vanishing  of  the  spatial components of the
electromagnetic current density $j^{k}$ in  the unitary gauge (and consequently
in  any  other   gauge,   since   the   electromagnetic   current   density  is
gauge-invariant).
Since in the unitary gauge $j^{k}=-2e^{2}A^{k}\text{Re}\left(\phi \right)$, the
spatial components $A^{k}$ of the electromagnetic potential also vanish.
Thus, we have the following  ansatz for the $Q$-ball fields in the unitary gauge
\begin{equation}
\phi \left(\mathbf{x},t\right) = \frac{f\left(r\right)}{\sqrt{2}},\;
A^{\mu}\left(\mathbf{x},t\right)=\eta^{\mu 0}A_{0}\left(r\right), \label{III:1}
\end{equation}
where  $f\left( r \right)$  and  $A_{0}\left( r \right)$  are  the  real ansatz
functions depending on the radial variable $r$.

Since the $Q$-ball is an unconditional extremum of the functional $F=E -\lambda
Q_{N}$, the first variation of $F$ vanishes in the neighborhood of the $Q$-ball
solution
\begin{equation}
\delta F = \delta E - \lambda \delta Q_{N} = 0.                   \label{III:2}
\end{equation}
Eq.~(\ref{III:2})  holds for arbitrary variations in fields in the neighborhood
of the $Q$-ball solution, including those that change  the $Q$-ball solution to
an infinitesimally close one.
It follows that the  energy  of  the  gauged  $Q$-ball  satisfies the important
relation
\begin{equation}
\frac{dE}{dQ_{N}} = \lambda,                                      \label{III:3}
\end{equation}
where the Lagrange multiplier  $\lambda$  is  a  function of the Noether charge
$Q_{N}$.
Since the $Q$-ball's energy and  the Noether  charge  are  gauge-invariant, the
Lagrange multiplier $\lambda$ is also gauge-invariant.
It is shown  in  Appendix A that in the unitary gauge,  the Lagrange multiplier
$\lambda$ is  expressed in terms of  the  limiting value of the electromagnetic
potential at spatial infinity:
\begin{equation}
\lambda = - e \underset{r\rightarrow \infty }{\lim } A_{0}(r)
\equiv \Omega_{\infty}.                                           \label{III:4}
\end{equation}

Substituting  ansatz  (\ref{III:1})  into  field  equations  (\ref{II:6a})  and
(\ref{II:6b}), we obtain  a  system of nonlinear differential equations for the
ansatz functions $f\left(r\right)$ and $\Omega\left( r \right) = -e A_{0}\left(
r \right)$:
\begin{eqnarray}
& & f^{\prime \prime}\left(r\right) + \frac{2}{r}f^{\prime}\left( r \right)
-\left( m^{2}-\Omega (r)^{2}\right) f\left( r\right)             \nonumber
\\
& & + \frac{g}{2}f\left(r\right)^{3} - \frac{h}{4}f\left(r\right)^{5} = 0,
                                                                 \label{III:5a}
\\
& & \Omega^{\prime \prime }(r)+\frac{2}{r}\Omega ^{\prime }(r)-
e^{2}\Omega(r)f\left( r\right)^{2} = 0.                          \label{III:5b}
\end{eqnarray}
The regularity of the $Q$-ball field configuration  and  the  finiteness of the
$Q$-ball's energy lead to  the  boundary  conditions  for the ansatz functions:
\begin{eqnarray}
&&f^{\prime }\left( 0\right) =0,\quad f\left( r \right) \underset{r\rightarrow
\infty }{\longrightarrow }0,    \nonumber
\\
&&\Omega ^{\prime }\left( 0\right) =0,\quad
\Omega \left( r\right)\underset{r\rightarrow \infty }{\longrightarrow}
\Omega_{\infty}.                                                  \label{III:6}
\end{eqnarray}

The general  properties  of the electromagnetic potential $A_{0}$ of the gauged
$Q$-ball were established in Ref.~\cite{klee}.
In terms of the ansatz function $\Omega(r)$, these properties are formulated as
\begin{subequations} \label{III:7}
\begin{eqnarray}
0 &<&\Omega \left( r_{1} \right) < \Omega \left( r_{2}\right) < \Omega_{\infty}
\quad \text{if} \quad \Omega_{\infty} > 0,                       \label{III:7a}
\\
0 &>&\Omega \left( r_{1}\right) > \Omega \left( r_{2}\right) >  \Omega_{\infty}
\quad \text{if} \quad \Omega_{\infty} < 0,                       \label{III:7b}
\end{eqnarray}
\end{subequations}
where it is supposed that $r_{1} < r_{2}$.
We see that $\Omega\left( r \right)$ is a positive and increasing (negative and
decreasing) function of $r$ if its limiting value $\Omega_{\infty}$ is positive
(negative).

The Lagrangian (\ref{II:1}) is invariant under the charge conjugation: $A_{\mu}
\rightarrow -A_{\mu },\,\phi \rightarrow \phi^{\ast}$.
This   invariance   of   the   Lagrangian   results   in   the   invariance  of
Eqs.~(\ref{III:5a})   and   (\ref{III:5b})  under  the  change of sign  $\Omega
\rightarrow -\Omega$.
Furthermore, the invariance  of  the  Lagrangian  (\ref{II:1})  under the other
discrete transformation $\phi \rightarrow -\phi$ (which is a particular case of
the phase transformation $\phi \rightarrow \exp \left( -i \alpha \right) \phi$)
leads to the invariance of Eqs.~(\ref{III:5a})   and   (\ref{III:5b}) under the
change of sign $f \rightarrow -f$.

We now consider  the  asymptotic  behavior  of  the gauged $Q$-ball solution at
small and large $r$.
Substituting   the   power   expansions   for   the   ansatz   functions   into
Eqs.~(\ref{III:5a}) and (\ref{III:5b}), we  obtain  the  asymptotic form of the
$Q$-ball solution at small $r$
\begin{subequations} \label{III:8}
\begin{eqnarray}
f(r) &=&f_{0}+\frac{f_{2}}{2!}r^{2}+O\left( r^{2}\right),        \label{III:8a}
\\
\Omega \left( r\right)  &=&\Omega_{0}+\frac{\Omega _{2}}{2!}r^{2}+
O\left(r^{2}\right),                                             \label{III:8b}
\end{eqnarray}
\end{subequations}
where the next-to-leading coefficients are  expressed  in  terms of $f_{0}$ and
$\Omega_{0}$
\begin{subequations} \label{III:9}
\begin{eqnarray}
f_{2} &=&\frac{1}{3}\left(\left(\Omega_{0}^{2} - m^{2}\right)
f_{0}-\frac{g}{2}f_{0}^{3}+\frac{h}{4}f_{0}^{5}\right),          \label{III:9a}
\\
\Omega_{2} &=&\frac{1}{3}e^{2}f_{0}^{2}\Omega_{0}.               \label{III:9b}
\end{eqnarray}
\end{subequations}
At large $r$, Eq.~(\ref{III:5b}) can be linearized and we obtain the asymptotic
form of $\Omega\left( r \right)$ as $r \rightarrow \infty$:
\begin{equation}
\Omega\left( r \right)  \sim \Omega_{\infty}
- \frac{e}{4\pi }\frac{Q}{r},                                    \label{III:10}
\end{equation}
where $Q = 4 \pi \int\nolimits_{0}^{\infty } j_{0}\left(r\right)r^{2}dr$ is the
total electric charge of the gauged $Q$-ball.
We see that due to the long-range character of the electromagnetic interaction,
the ansatz  function  $\Omega$  tends  rather  slowly  ($\sim r^{-1}$)  to  the
limiting value $\Omega_{\infty}$.
Due to this, in Eq.~(\ref{III:5a}), the  gauge field does not decouple from the
scalar field  even at large $r$.
This results in  the  following asymptotics \cite{gulamov_2015} for the complex
scalar field as $r \rightarrow \infty$
\begin{equation}
f\left( r\right)  \sim f_{\infty}\left(\Delta r\right)^{-\left(1 + \frac{\beta}
{2\Delta }\right) }\exp \left( -\Delta r\right),                 \label{III:11}
\end{equation}
where
\begin{equation}
\Delta = \left( m^{2}-\Omega_{\infty}^{2}\right) ^{1/2},\quad
\beta = \frac{e}{2\pi}\Omega_{\infty} Q,                         \label{III:12}
\end{equation}
and $f_{\infty}$ is a constant.
We see that the long-range tail of the gauge field leads to  a  faster decrease
in the complex scalar field of the  gauged  $Q$-ball at large $r$ in comparison
with the nongauged $Q$-ball for which $f\left( r \right) \sim f_{\infty} \left(
\Delta r \right)^{-1}\exp \left(- \Delta r \right)$.
As  $\left\vert\Omega_{\infty}\right\vert=m$, the exponent $ - \left(1 + 2^{-1}
\beta \Delta ^{-1}  \right)$ of   the    preexponential   factor  diverges, and
Eq.~(\ref{III:11}) becomes inapplicable.
However,  it   was  shown  in  Ref.~\cite{gulamov_2015}  that  when $\left\vert
\Omega_{\infty} \right\vert = m$, the scalar  field  of the gauged $Q$-ball has
the following asymptotics at large $r$
\begin{equation}
f\left( r\right) \sim f_{\infty}\left(m r\right)^{-3/4}
\exp\left(-\sqrt{\frac{2e}{\pi}Q m r}\right).                    \label{III:13}
\end{equation}
Thus, unlike the nongauged  $Q$-ball,  the  gauged  $Q$-ball also exists at the
limiting point $\left\vert \Omega_{\infty}  \right\vert = m$  and  has a finite
energy and Noether charge at this point.
At the  same  time,  like the nongauged $Q$-ball, the  gauged $Q$-ball does not
exist when  $\left\vert  \Omega_{\infty} \right\vert > m$ since the asymptotics
(\ref{III:11}) shows oscillating  behaviour, leading  to an infinite energy and
Noether charge for the corresponding field configuration.

The electromagnetic current density  and  the components of the energy-momentum
tensor can also be expressed in terms of the ansatz functions:
\begin{equation}
j_{\mu} = e \Omega(r)f\left( r\right)^{2}\eta_{\mu 0},           \label{III:14}
\end{equation}
\begin{eqnarray}
T_{00} &=&\frac{1}{2e^{2}}\Omega^{\prime }\left( r\right)^{2}
+ \frac{1}{2} f^{\prime} \left( r \right)^{2}                   \label{III:15a}
\\
& & +\frac{1}{2} \Omega(r)^{2}f\left( r\right)^{2}
+ V\left( f\left( r \right) \right),                             \nonumber
\\
T_{0k} &=&0,                                                    \label{III:15b}
\\
T_{ij} &=&\left( \frac{x_{i}x_{j}}{r^{2}}-\frac{1}{3}\delta _{ij}\right)
s\left( r\right) +\delta _{ij}p\left( r\right),                 \label{III:15c}
\end{eqnarray}
where the radially dependent functions
\begin{equation}
s\left( r\right) = f^{\prime }\left( r\right) ^{2}-e^{-2}\Omega^{\prime
}\left( r\right)^{2},                                           \label{III:16a}
\end{equation}
and
\begin{eqnarray}
p\left( r\right)&=& \frac{1}{6e^{2}}\Omega ^{\prime }\left( r\right)^{2}-
\frac{1}{6}f^{\prime }\left( r\right) ^{2} \nonumber \\
&&+\frac{1}{2}\Omega \left( r\right)^{2}f\left( r\right)^{2}
-V\left(f\left( r\right)\right)                                 \label{III:16b}
\end{eqnarray}
are the shear force  and  pressure  distributions, respectively.
From Eqs.~(\ref{III:7}) and (\ref{III:14}), it follows that
\begin{equation}
\text{sign}\left( j_{0}\left( r\right) \right) = \text{sign}\left( \Omega
\left(r\right) \right)=\text{sign}\left(\Omega_{\infty}\right),  \label{III:17}
\end{equation}
and hence the sign of the electric (Noether) charge coincides  with that of the
parameter $\Omega_{\infty}$.
In contrast, it follows  from  Eqs.~(\ref{III:15a}) -- (\ref{III:16b}) that the
values of the components of  the  energy-momentum  tensor  do not depend on the
sign of $\Omega_{\infty}$.
We conclude  that the energy  of  the  gauged  $Q$-ball is  an even function of
$\Omega_{\infty}$, whereas  the electric  (Noether)  charge is an odd function:
\begin{equation}
E\left(-\Omega_{\infty}\right) =   E\left(\Omega_{\infty}\right) ,\;
Q\left(-\Omega_{\infty}\right) = - Q\left(\Omega_{\infty}\right).\label{III:18}
\end{equation}

Since the $T_{0k}$ components of the energy-momentum tensor vanish, the angular
momentum of the spherically symmetrical gauge $Q$-ball  is  equal  to  zero, as
expected.
The conservation  of  the  energy-momentum  tensor  leads  to  the differential
relation between the shear force  and the pressure
\begin{equation}
\frac{2}{r}s\left( r\right) +\frac{2}{3}s^{\prime }\left( r \right)
+ p^{\prime}\left( r\right) = 0.                                 \label{III:19}
\end{equation}
Multiplying  Eq.~(\ref{III:19})  by  $r^{3}$  and integrating by parts over $r$
from zero to  infinity,  we  obtain  the Laue condition \cite{Laue, birula} for
the pressure distribution
\begin{equation}
\int\nolimits_{0}^{\infty }drr^{2}p\left( r\right) = 0.          \label{III:20}
\end{equation}

Any solution of field  equations (\ref{II:6a}) and (\ref{II:6b}) is an extremum
of the action $S = \int \mathcal{L}d^{3}xdt$.
In the case of the gauged $Q$-ball, the Lagrangian  density  $\mathcal{L}$ does
not depend on time, and thus the  gauged  $Q$-ball  solution is  an extremum of
the Lagrangian $L = \int\mathcal{L}d^{3}x$.
Next, for the $Q$-ball solution, the total energy $E=\int T_{00}d^{3}x$ and the
Lagrangian $L = \int \mathcal{L}d^{3}x$ can be presented as linear combinations
of the  electrostatic,  gradient,  kinetic,  and  potential terms:
\begin{eqnarray}
E &=&E^{\left( E\right) }+E^{\left( G\right) } + E^{\left( T\right)
}+E^{\left( P\right) },                                         \label{III:21a}
\\
L &=&E^{\left( E\right) }-E^{\left( G\right)}+E^{\left( T\right)}
-E^{\left( P\right) },                                          \label{III:21b}
\end{eqnarray}
where
\begin{subequations} \label{III:22}
\begin{eqnarray}
E^{\left( E\right) } &=&4\pi \int\nolimits_{0}^{\infty }\frac{1}{2e^{2}}
\Omega ^{\prime }\left( r\right) ^{2}r^{2}dr,                   \label{III:22a}
\\
E^{\left( G\right) } &=&4\pi \int\nolimits_{0}^{\infty }\frac{1}{2}
f^{\prime }\left( r\right) ^{2}r^{2}dr,                         \label{III:22b}
\\
E^{\left( T\right) } &=&4\pi \int\nolimits_{0}^{\infty }\frac{1}{2}\Omega
(r)^{2}f\left( r\right) ^{2}r^{2}dr,                            \label{III:22c}
\\
E^{\left( P\right) } &=&4\pi \int\nolimits_{0}^{\infty }V\left( f\left(
r\right) \right) r^{2}dr.                                       \label{III:22d}
\end{eqnarray}
\end{subequations}
Let us  consider  the  scale  transformation  of  the  gauge $Q$-ball's fields:
$f\left( r\right) \rightarrow f \left( \kappa r\right) ,\, A_{0}\left(r \right)
\rightarrow A_{0}\left( \kappa r \right)$.
Under this transformation, the terms in Eqs.~(\ref{III:22a}) -- (\ref{III:22d})
behave as follows:  $E^{\left( E\right)}   \rightarrow   \kappa^{-1}E^{\left( E
\right)}$,  $E^{\left( G \right)} \rightarrow \kappa^{-1}E^{\left( G \right)}$,
$E^{\left( T \right)}\rightarrow \kappa^{-3} E^{ \left( T \right) }$, and  $E^{
\left( P \right)}  \rightarrow  \kappa^{-3} E^{\left( P \right)}$, and thus the
Lagrangian (\ref{III:21b}) becomes  a function of the scale parameter $\kappa$.
Since  $f\left(\kappa r \right)$  and  $A_{0}\left(\kappa r \right)$  form  the
$Q$-ball solution  at $\kappa = 1$, the  derivative $dL/d\kappa$ must vanish at
this point: $\left. dL/d\kappa \right\vert_{\kappa = 1} = 0$.
The last  equation  results  in  the  virial  relation  for the gauged $Q$-ball
\begin{equation}
E^{\left( E\right) }-E^{\left( G\right) }+3\left( E^{\left( T\right)
}-E^{\left( P\right) }\right) = 0.                               \label{III:23}
\end{equation}
It can be  easily  shown  that  given  Eqs.~(\ref{III:22a}) -- (\ref{III:22d}),
the  virial  relation  (\ref{III:23})  is  equivalent  to  the  Laue  condition
(\ref{III:20}).

The energy of the gauged $Q$-ball can be presented in several equivalent forms.
Integrating the term $\Omega^{\prime 2}$  in  the  electrostatic energy density
$\Omega^{\prime}\left(r\right)^{2}/(2e^{2})$ by parts, using Eq.~(\ref{III:5b})
(Gauss's  law), and taking into account  the boundary conditions (\ref{III:6}),
we obtain the following  expression  for the electrostatic energy of the gauged
$Q$-ball
\begin{eqnarray}
E^{\left( E\right) } &=&\frac{1}{2}4\pi \int\limits_{0}^{\infty }\left(
A_{0}\left( r \right) -  A_{0}\left( \infty \right) \right) j_{0}\left(
r\right) r^{2}dr                                              \nonumber
\\
&=&\frac{1}{2e}4\pi \int\limits_{0}^{\infty }\left( \Omega _{\infty
}-\Omega \left(r\right) \right) j_{0}\left(r\right) r^{2}dr.    \label{III:23a}
\end{eqnarray}
The kinetic energy  (\ref{III:22c})  can  also  be  rewritten  in  terms of the
electric charge density:
\begin{eqnarray}
E^{\left( T\right) } &=&-\frac{1}{2}4\pi \int\limits_{0}^{\infty
}A_{0}\left( r\right) j_{0}\left( r\right) r^{2}dr      \nonumber
\\
&=&\frac{1}{2e}4\pi \int\limits_{0}^{\infty}\Omega \left(r\right)
j_{0}\left( r\right) r^{2}dr.                                   \label{III:23b}
\end{eqnarray}
Combining      Eqs.~(\ref{III:21a}),      (\ref{III:22b}),     (\ref{III:22d}),
(\ref{III:23a}), and (\ref{III:23b}) results in an  alternative  expression for
the energy of the gauged $Q$-ball
\begin{eqnarray}
E &=&\Omega_{\infty} \frac{Q_{N}}{2} + 4 \pi \int\nolimits_{0}^{\infty}
\left(\frac{1}{2} f^{\prime}\left( r\right)^{2} +
V\left( f\left( r \right) \right) \right) r^{2}dr
\nonumber
\\
&=&\Omega_{\infty} \frac{Q_{N}}{2}+E^{\left( G\right) }+E^{\left(P\right)}.
                                                                 \label{III:24}
\end{eqnarray}
Next, Eqs.~(\ref{III:21a}) and (\ref{III:24}) lead to the  following expression
for the Noether charge
\begin{equation}
Q_{N}=\frac{2}{\Omega_{\infty}}\left(E^{\left( E\right)} +
E^{\left(T\right)}\right).                                       \label{III:25}
\end{equation}
Finally,  using  Eqs.~(\ref{III:24}),   (\ref{III:25}),  and the virial relation
(\ref{III:23}), we obtain  two  more  expressions  for  the energy of the gauged
$Q$-ball:
\begin{eqnarray}
E & = &\Omega_{\infty} Q_{N}+\frac{8\pi }{3}\int\nolimits_{0}^{\infty }
\left(\frac{1}{2}f^{\prime}\left(r\right)^{2}-
\frac{1}{2 e^{2}}\Omega^{\prime}\left( r\right)^{2}\right) r^{2}dr \nonumber
\\
&=&\Omega_{\infty} Q_{N}+\frac{2}{3}\left( E^{\left( G\right) } -
E^{\left( E\right)}\right)                                       \label{III:26}
\end{eqnarray}
and
\begin{eqnarray}
E &=&\Omega_{\infty} Q_{N}+8\pi \int\nolimits_{0}^{\infty }
\left(\frac{1}{2}\Omega(r)^{2}f\left( r\right) ^{2} -
V\left( f\left( r\right) \right) \right) r^{2}dr                 \nonumber
\\
&=&\Omega_{\infty} Q_{N}+2\left( E^{\left( T\right)}
- E^{\left( P\right) }\right).                                   \label{III:27}
\end{eqnarray}

\section{\label{sec:IV} Numerical results}

The system of  differential  equations  (\ref{III:5a}) and  (\ref{III:5b}) with
boundary conditions (\ref{III:6}) represents a  mixed boundary value problem on
the semi-infinite interval $r \in \left[0, \infty \right)$, which can be solved
only by numerical methods.
In this paper, the boundary value problem  was  solved  using  the {\sc{Maple}}
package \cite{maple}.
Since the  point  $r  =  0$  is  the  regular  singular  point  of  the  system
(\ref{III:5a}), (\ref{III:5b}), we apply a difference  scheme that does not use
the boundary values of the functions.
To check the correctness  of our  numerical solutions, we use Eq.~(\ref{III:3})
and the Laue condition (\ref{III:20}).

The mixed boundary value problem (\ref{III:5a}) -- (\ref{III:6}) depends on the
five parameters: $e$, $m$, $g$, $h$, and $\Omega_{\infty}$.
To reduce the number of the parameters,  we rescale the radial variable and the
ansatz functions:
\begin{equation}
r=\frac{\tilde{r} }{m},\;
f\left( r\right)=\frac{m}{\sqrt{g}}\tilde{f}\left(\tilde{r} \right),\;
A_{0}\left(r\right)=\frac{m}{\sqrt{g}}\tilde{A}_{0}\left(\tilde{r}
\right).                                                           \label{IV:1}
\end{equation}
After rescaling,  the  boundary  value  problem  will depend  on only the three
dimensionless parameters: $\tilde{e} = e g^{-1/2}$, $\tilde{h} = hm^{2}g^{-2}$,
and $\tilde{\Omega}_{\infty} = m^{-1}\Omega_{\infty}$.
In  particular,  the   self-interaction   potential  will  depend  on  only one
dimensionless parameter: $V(\tilde{f})=2^{-1}\tilde{f}^{2}- 8^{-1}\tilde{f}^{4}
+ 24^{-1}\tilde{h}\tilde{f}^{6}$.
Furthermore, the dependences of the Noether charge $Q_{N}$, the electric charge
$Q$, and the energy $E$ on the parameters $m$ and $g$ are factorized as follows:
\begin{equation}
Q_{N} = g^{-1} \tilde{Q}_{N},\quad
Q = g^{-1/2} \tilde{Q},\quad
E = m g^{-1} \tilde{E},                                            \label{IV:2}
\end{equation}
where $\tilde{Q}_{N}$, $\tilde{Q}=\tilde{e}\tilde{Q}_{N}$,  and $\tilde{E}$ are
the  rescaled  versions  of  the Noether  charge, the electric  charge, and the
energy, respectively.
We can  therefore  without  loss  of  generality set the parameters $m$ and $g$
equal to unity.
To avoid spontaneous breaking of the gauge symmetry, the sextic self-interaction
coupling constant  $\tilde{h}$  must satisfy the inequality $\tilde{h} > 3/16$.
In most numerical calculations, we set $\tilde{h} = 0.2$, whereas the remaining
parameters $\tilde{e}$ and $\tilde{\Omega}_{\infty}$  may vary within allowable
intervals.

\begin{figure}[t]
\includegraphics[width=0.5\textwidth]{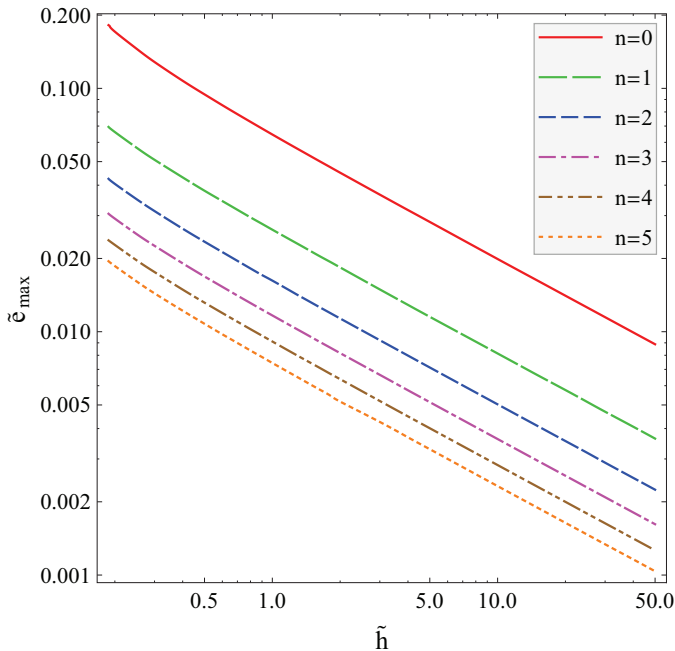}
\caption{\label{fig1}  Dependences  of  the  maximum  allowable  gauge coupling
constant $\tilde{e}_{\text{max}}$  on  the  self-interaction  coupling constant
$\tilde{h}$ for the basic and first five radially excited $Q$-ball states.}
\end{figure}

First, let us ascertain  the  domain in the  parameter space in which the basic
and radially excited states of the gauged $Q$-ball can exist.
It is shown in Appendix B that  the  gauged  $Q$-ball  solution cannot exist if
the gauge coupling constant exceeds some maximum value.
Figure~\ref{fig1} shows   the   dependences  of  the  maximum  allowable  gauge
coupling constant $\tilde{e}_{\max}$ on  the self-interaction coupling constant
$\tilde{h}$  for  the  basic  and  the  first five  radially excited $Q$-balls.
These dependences  are  presented  on  the  logarithmic  scale ranging from the
minimum permissible $\tilde{h} = 3/16$ to $\tilde{h} = 50$.
We see that for $\tilde{h} \gtrsim 0.5$, all  the curves  are well described by
the formula
\begin{equation}
\tilde{e}_{\max } \approx \frac{\epsilon_{n}}{\sqrt{\tilde{h}}},   \label{IV:3}
\end{equation}
where $\epsilon_{n}$ are constant coefficients and the label $n = 0, \ldots, 5$
corresponds to the basic and first five radially excited $Q$-balls.

For large $\tilde{h}$, Eq.~(\ref{IV:3}) can  be explained as follows.
Let $\tilde{e}$ be equal to zero; then the electromagnetic field decouples from
the complex scalar field and the  ansatz  function $\tilde{\Omega}$ becomes the
constant phase frequency $\tilde{\omega}$.
In this nongauged case, there  is  an  analytical  expression  for  the minimum
possible  phase  frequency: $\tilde{\omega}_{\min}  =  (1 - \left( 3/16 \right)
\tilde{h}^{-1})^{1/2} = 1-\left(3/32\right)\tilde{h}^{-1} + O(\tilde{h}^{-2})$.
We see that for sufficiently  large  $\tilde{h}$,  the phase frequency $\tilde{
\omega}$  lies  in  the narrow range $( 1-\left(3/32\right)\tilde{h}^{-1},\,1)$
with width $\propto \tilde{h}^{-1}$.
Next, we turn on the electromagnetic  interaction by allowing $\tilde{e}$ to be
different from zero,  meaning that  the  phase frequency $\tilde{\omega}$ turns
into the monotonically increasing  (see Eq.~(\ref{III:7a}))  function  $\tilde{
\Omega}(\tilde{r})$.
This increase in $\tilde{e}$ leads to an increase in $\tilde{\Omega}_{\infty}$,
which continues  until $\tilde{\Omega}_{\infty}$ reaches  the  maximum possible
value $\tilde{\Omega}_{\infty} = 1$ at $\tilde{e} = \tilde{e}_{\max}$.
At the same time, Eqs.~(\ref{III:14})  and  (\ref{B:1})  tell us that for small
$\tilde{e}$,  the  difference  $\tilde{\Omega}_{\infty} - \tilde{\Omega}_{0}  =
\tilde{e}^{2}\tilde{J}_{N}+O(\tilde{e}^{4})$, where the integral $\tilde{J}_{N}
= \tilde{\omega} \int\nolimits_{0}^{\infty} \tilde{r} \tilde{f} \left(\tilde{r}
\right)^{2}d\tilde{r}$ does not depend on $\tilde{e}$.
It follows that  $\tilde{e}^{2}_{\max} \propto \tilde{h}^{-1}$  at sufficiently
large $\tilde{h}$, resulting in Eq.~(\ref{IV:3}).
Note, however, that Eq.~(\ref{IV:3}) becomes  valid at $\tilde{h} \gtrsim 0.5$.

From Fig.~\ref{fig1}, it  follows  that  in  Eq.~(\ref{IV:3}), the coefficients
$\epsilon_{n}$ decrease with an increase in $n$.
Thus,  the   maximum  allowable  gauge  coupling   constant  $\tilde{e}_{\max}$
becomes smaller for the more excited $Q$-ball solutions.
It was found numerically that for $n\gtrsim 3$, the coefficients $\epsilon_{n}$
are well described by the formula
\begin{equation}
\epsilon_{n} \approx 0.037 \, n^{-1}.                              \label{IV:4}
\end{equation}
We see that if the  point $(\tilde{h},\tilde{e})$ lies above the solid curve in
Fig.~\ref{fig1}, then no $Q$-ball solutions exist with the parameters $\tilde{h
}$ and $\tilde{e}$.
Moreover, if the gauge coupling constant $\tilde{e}$ exceeds the limiting value
$ 0.182$, then there are no $Q$-ball solutions in model (\ref{II:1}).
On the other hand, Eqs.~(\ref{IV:3}) and (\ref{IV:4}) tell us that  for a given
$\tilde{h}$, the number of  radially  excited  states of the gauged $Q$-ball is
inversely proportional to the gauge coupling constant $\tilde{e}$.
Thus, the  number  of  radially excited states increases indefinitely ($\propto
\tilde{e}^{-1}$) as the gauge coupling constant $\tilde{e} \rightarrow 0$.

\begin{figure}[t]
\includegraphics[width=0.5\textwidth]{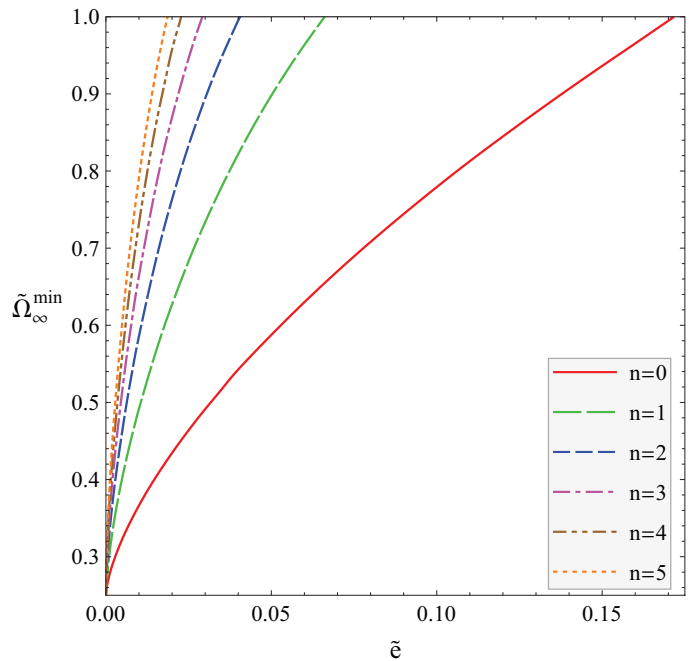}
\caption{\label{fig2}    Dependences  of  the  minimum  allowable  value of the
parameter $\tilde{\Omega}_{\infty}$ on the gauge coupling  constant $\tilde{e}$
for the basic  and  first  five  radially  excited $Q$-ball states.  The curves
correspond  to  the  self-interaction  coupling  constant $\tilde{h} = 0.2$.}
\end{figure}

Let  a  $Q$-ball solution  exist for given values of the parameters $\tilde{e}$
and $\tilde{h}$.
In this case,  the  parameter $\tilde{\Omega}_{\infty}$  lies in  a  range from
the minimum allowable  value  $\tilde{\Omega}_{\infty}^{\min}$ to  the  maximum
allowable  value of $1$.
Figure~\ref{fig2} presents the  dependences  of  the minimum allowable value of
$\tilde{\Omega}_{\infty}$ on the  gauge  coupling  constant $\tilde{e}$ for the
basic and the first five radially excited $Q$-ball solutions.
In   Fig.~\ref{fig2},  the  curves  $\tilde{\Omega}_{\infty}^{\min}(\tilde{e})$
correspond  to   the   self-interaction  coupling  constant  $\tilde{h} = 0.2$;
for other values of $\tilde{h}$,  the behaviour of the curves $\tilde{\Omega}_{
\infty}^{\min}(\tilde{e})$ is similar to that in Fig.~\ref{fig2}.
We see  that  for  a  given  $\tilde{e}$,  the $n$-th radially excited $Q$-ball
solution may exist only if $\tilde{\Omega}_{\infty} \in [\tilde{\Omega}_{\infty
}^{\min}(\tilde{e}), 1]$, where it  is  understood that $\tilde{\Omega}_{\infty
}^{\min}(\tilde{e})$ is taken for the corresponding $n$.
When $\tilde{e}$ tends to zero,  all the  curves in Fig.~\ref{fig2} tend to the
minimum possible phase frequency for the nongauged case: $\tilde{\omega}_{\min}
= (1-\left(3/16\right) \tilde{h}^{-1})^{1/2}$.
As $\tilde{e}$  increases,  all  the  curves  monotonically  increase until the
maximum possible value $\tilde{\Omega}_{\infty} = 1$ is reached.
In Fig.~\ref{fig2},  the  intersection  points  of  the  curves  and  the  line
$\tilde{\Omega}_{\infty} = 1$  represent  the  maximum allowable gauge coupling
constants corresponding  to the  self-interaction constant $\tilde{h} = 0.2$ in
Fig.~\ref{fig1}.
In  particular, it follows from   Fig.~\ref{fig2}    that  in  accordance  with
Fig.~\ref{fig1}, the maximum allowable gauge coupling constant $\tilde{e}_{\max
}$ decreases with an increase in the radial excitation of the $Q$-ball solution.

\begin{figure}[t]
\includegraphics[width=0.5\textwidth]{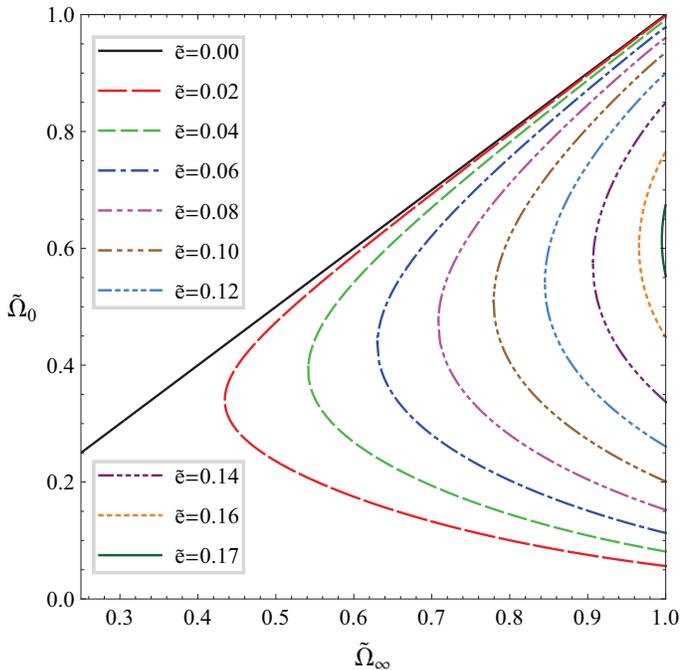}
\caption{\label{fig3}      Curves $\tilde{\Omega}_{0}(\tilde{\Omega}_{\infty})$
for  different  values  of  the  gauge  coupling  constant  $\tilde{e}$ and the
self-interaction coupling constant $\tilde{h} = 0.2$.  The curves correspond to
the  unexcited  $Q$-ball  solution. The straight solid line  corresponds to the
linear  dependence $\tilde{\Omega}_{0} = \tilde{\Omega}_{\infty}$.}
\end{figure}

\begin{figure}[t]
\includegraphics[width=0.5\textwidth]{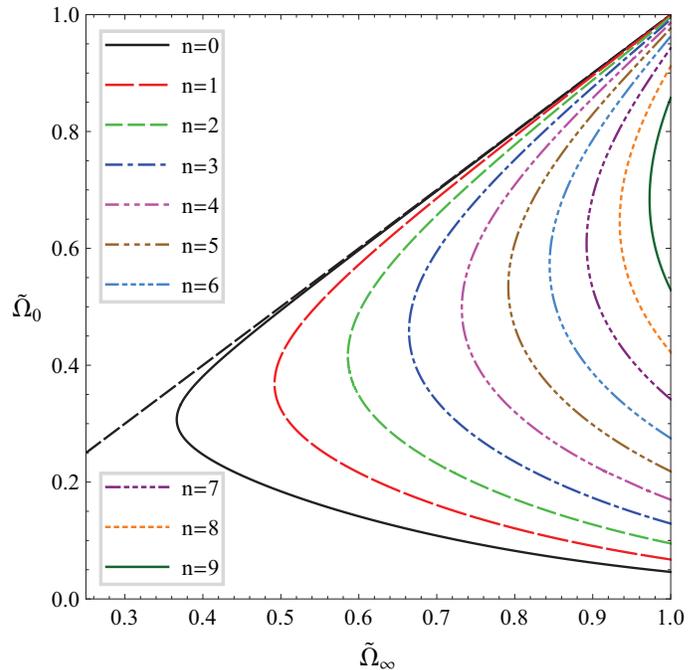}
\caption{\label{fig4}     Curves  $\tilde{\Omega}_{0}(\tilde{\Omega}_{\infty})$
for  the  unexcited  and  first  nine  radially   excited  $Q$-ball  solutions.
The curves correspond to the parameters $\tilde{e}=0.01$ and $\tilde{h} = 0.2$.
The straight dashed line  corresponds to the  linear  dependence $\tilde{\Omega
}_{0} = \tilde{\Omega}_{\infty}$.}
\end{figure}

It is  shown  in  Appendix  B  that  the  restriction $\Delta\Omega < m$ on the
difference $\Delta\Omega = \Omega_{\infty} - \Omega_{0}$ prevents the existence
of gauged $Q$-balls with an arbitrarily large electric charge or gauge coupling
constant.
However, the parameters $\Omega_{0}$ and $\Omega_{\infty}$ are not independent,
since they are both determined  by  the  solution  to  the mixed boundary value
problem (\ref{III:5a}) -- (\ref{III:6}).
In this connection, it  would  be  interesting to study the dependence $\tilde{
\Omega}_{0}( \tilde{\Omega}_{\infty} )$  for  different  values  of  the  gauge
coupling constant as well as for different radially excited $Q$-ball solutions.
Figure~\ref{fig3} shows  the  dependences  $\tilde{\Omega}_{0}(\tilde{\Omega}_{
\infty})$ for the unexcited $Q$-ball solution corresponding to different values
of the gauge coupling constant $\tilde{e}$.
We see that for non-zero $\tilde{e}$, all the curves $\tilde{\Omega}_{0}(\tilde{
\Omega}_{\infty})$ intersect  the  line  $\tilde{\Omega}_{\infty} = 1$  at  two
points, and thus each of these curves  has a turning point at $\tilde{\Omega}_{
\infty}  =  \tilde{\Omega}_{\infty}^{\min}$,  where  the  derivative  $d\tilde{
\Omega}_{0}/d\tilde{\Omega}_{\infty}$ becomes infinite.
In  Fig.~\ref{fig3},  the  dependence  of  $\tilde{\Omega}_{\infty}^{\min}$  on
$\tilde{e}$ is described by the solid curve in Fig.~\ref{fig2}.
In particular, it follows from Fig.~\ref{fig2} that $1 - \tilde{\Omega}_{\infty
}^{\min }\propto \tilde{e}_{\max } - \tilde{e}$  in the vicinity of the maximum
allowable gauge coupling constant $\tilde{e}_{\max}$.
This means that the  allowable interval of $\tilde{\Omega}_{\infty}$ shrinks to
a point as $\tilde{e}\rightarrow \tilde{e}_{\max}$ and thus the gauged $Q$-ball
cannot exist if $\tilde{e} \ge \tilde{e}_{\max}$.
From Fig.~\ref{fig3}, it follows  that  the parameter $\tilde{\Omega}_{\infty}$
does not  uniquely define the  gauged  $Q$-ball solution, since there exist two
different  $Q$-ball  solutions  for  each  $\tilde{\Omega}_{\infty}\in (\tilde{
\Omega}_{\infty}^{\min}, 1]$.
At the same time, Fig.~\ref{fig3} tells us that the gauged $Q$-ball solution is
uniquely defined by the parameter $\tilde{\Omega}_{0}$.
The curves shown in Fig.~\ref{fig3} correspond to the  basic ($n = 0$) $Q$-ball
solution; for the radially excited  $Q$-ball solutions ($n \ge 1$) the behavior
of the  curves $\tilde{\Omega}_{0}(\tilde{\Omega}_{\infty})$ is similar to that
in Fig.~\ref{fig3}.

Now we discuss the behavior of  gauged $Q$-balls as the gauge coupling constant
$\tilde{e}$ tends to zero.
From Fig.~\ref{fig3}, it follows that turning points divide the curves $\tilde{
\Omega}_{0}(\tilde{\Omega}_{\infty})$ into upper and lower branches.
In the limit $\tilde{e}\rightarrow 0$, the $\tilde{\Omega}_{\infty}$ coordinate
of the turning point tends to the nongauged value $\tilde{\omega}_{\min} = (1 -
\left(3/16\right) \tilde{h}^{-1})^{1/2}$,  and  thus gauged  $Q$-ball solutions
that lie on the  upper  branch  tend  to  the  corresponding  nongauged ones on
the straight line $\tilde{\Omega}_{0} = \tilde{\Omega}_{\infty}$.
In contrast, gauged $Q$-ball solutions that lie on the lower branch do not tend
to  nongauged ones as $\tilde{e} \rightarrow 0$.
It can be said that as $\tilde{e}\rightarrow 0$, the gauge field decouples from
$Q$-ball solutions lying on the  upper  branch of the curve $\tilde{\Omega}_{0}
(\tilde{\Omega}_{\infty})$ but does not decouple from  those lying on the lower
branch.
Indeed, in this limit, the difference  $\Delta\tilde{\Omega} = \tilde{\Omega}_{
\infty} - \tilde{\Omega}_{0}  \rightarrow 0$  for  $Q$-ball  solutions  on  the
upper branch, whereas $\Delta\tilde{\Omega} \rightarrow \tilde{\Omega}_{\infty}
- \tilde{\omega}_{\min}$ for $Q$-ball solutions on the lower branch.
It follows that  for  $Q$-ball  solutions  that  lie  on  the lower branch, the
integral $\tilde{J} = \int_{0}^{\infty}\tilde{r} \tilde{j}_{0} \left( \tilde{r}
\right) d\tilde{r}$   on   the  right-hand  side  of  Eq.~(\ref{B:1}) increases
indefinitely as $\tilde{e} \rightarrow 0$.
Hence, the electric  charge  $\tilde{Q} = 4 \pi \int_{0}^{\infty }\tilde{r}^{2}
\tilde{j}_{0}   \left(  \tilde{r}  \right)  d\tilde{r}$,   the  Noether  charge
$\tilde{Q}_{N} = \tilde{e}^{-1} \tilde{Q}$, and the energy $\tilde{E}$ of these
$Q$-ball solutions also increase indefinitely in this limit.

Now let us  consider  the  curves $\tilde{\Omega}_{0}(\tilde{\Omega}_{\infty})$
for the radially excited $Q$-ball solutions.
Figure~\ref{fig4} shows these  curves for the unexcited ($n = 0$) and the first
nine radially excited  ($n = 1, \ldots, 9$)  gauged  $Q$-balls corresponding to
the parameters $\tilde{e} = 0.01$ and $\tilde{h} = 0.2$.
We see that in Fig.~\ref{fig4}, the behaviour of  the curves is similar to that
in Fig.~\ref{fig3}.
Indeed, all the curves in  Fig.~\ref{fig4} intersect the limiting line $\tilde{
\Omega}_{\infty}= 1$ at two points,  and have a turning point with the infinite
derivative $d\tilde{\Omega}_{0}/d\tilde{\Omega}_{\infty}$.
This similarity can be explained as follows.
Eqs.~(\ref{IV:3}) and (\ref{IV:4}) tell us that for the $n$-th radially excited
$Q$-ball  solution, the  maximum  allowable  gauge  coupling constant $\tilde{e
}_{\max} \propto n^{-1}$. 
It follows that for the  curves in Fig.~\ref{fig4}, the difference $\tilde{e}_{
\max} - \tilde{e}$  decreases  with  an  increase in $n$ due to the decrease in
$\tilde{e}_{\max}$.
A similar  situation  is  observed  in  Fig.~\ref{fig3},  where  the difference
$\tilde{e}_{\max} - \tilde{e}$  decreases due to the increase in $\tilde{e}$ at
fixed $\tilde{e}_{\max}$.
In both cases, the  decrease  in  the difference $\tilde{e}_{\max} - \tilde{e}$
leads to a  decrease in  the  allowable  interval of $\tilde{\Omega}_{\infty}$.
The decrease  in  the  interval  of  $\tilde{\Omega}_{\infty}$  results  in the
existence  of  the  maximum  radially  excited  $Q$-ball  solution  for a given
$\tilde{e}$ (the solution with $n = 9$ in Fig.~\ref{fig4}).
Note that for a given $\tilde{e}$, the existence of the maximum radially excited
gauged $Q$-ball is in accordance with Eqs.~(\ref{IV:3}) and (\ref{IV:4}).

\begin{figure}[t]
\includegraphics[width=0.5\textwidth]{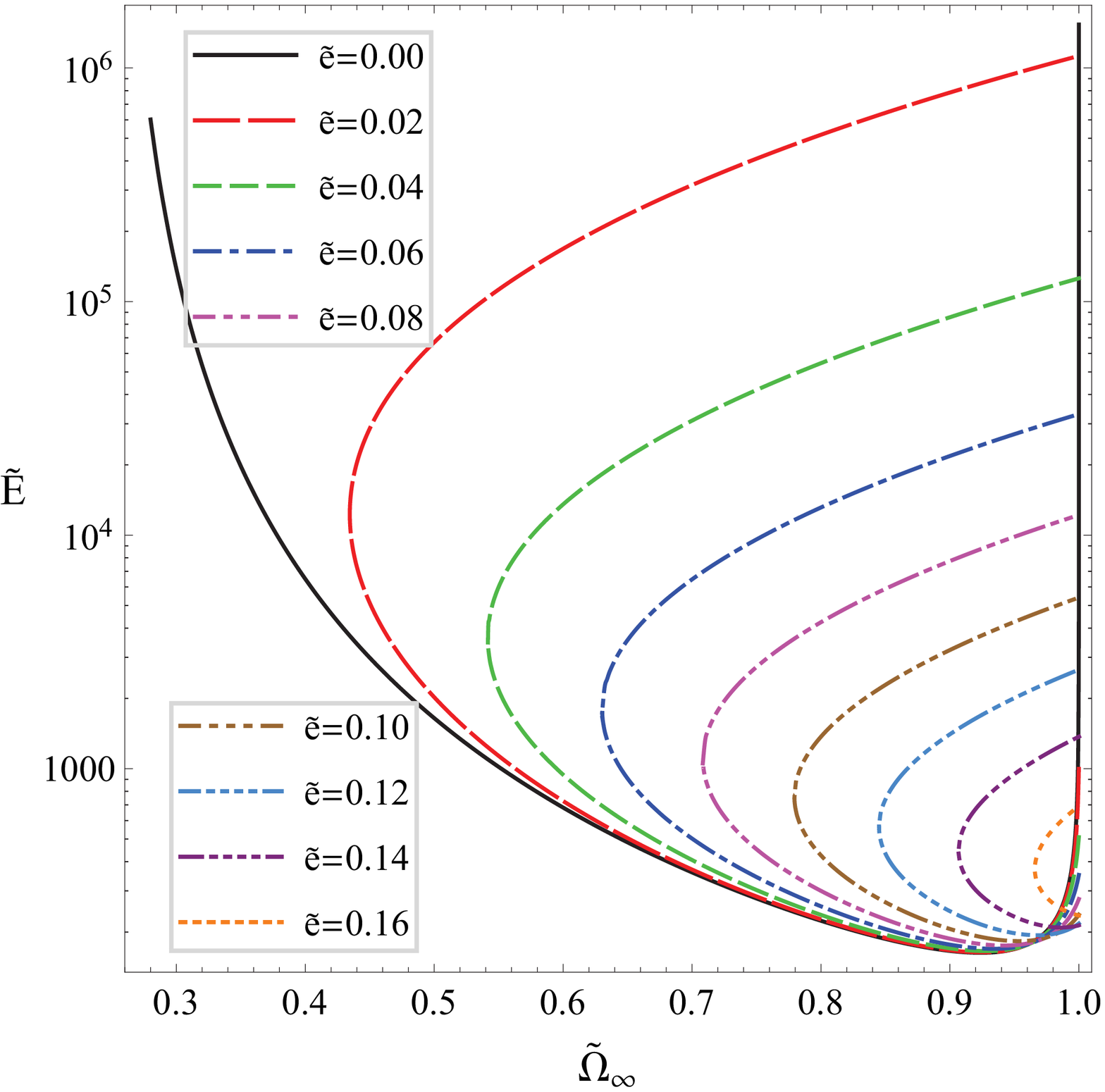}
\caption{\label{fig5} Curves $\tilde{E}(\tilde{\Omega}_{\infty})$ for different
values  of  $\tilde{e}$  and  $\tilde{h} = 0.2$.   The curves correspond to the
unexcited $Q$-ball solution.}
\end{figure}

Next, we turn to  a  study of the dependence  of  the energy $\tilde{E}$ on the
parameter $\tilde{\Omega}_{\infty}$ for the gauged $Q$-ball solutions.
The dependence $\tilde{Q}_{N}(\tilde{\Omega}_{\infty})$ is similar to $\tilde{E
}(\tilde{\Omega}_{\infty})$ and  is therefore not shown.
In Fig.~\ref{fig5}, we can see the  curves $\tilde{E}(\tilde{\Omega}_{\infty})$
for the different values of the gauge coupling constant $\tilde{e}$.
These curves correspond to the unexcited $Q$-ball solution.
The curves $\tilde{E}(\tilde{\Omega}_{\infty})$ corresponding to  the  radially
excited $Q$-ball solutions are  similar  to  those  in Fig.~\ref{fig5}, and are
not shown.
Firstly, we  see  that  the  behavior  of the curve $\tilde{E}(\tilde{\omega})$
for the nongauged  case $\tilde{e} = 0$ (for which the parameter $\tilde{\Omega
}_{\infty}$ is  equal  to  the phase frequency $\tilde{\omega}$) is drastically
different from that for the gauged case $\tilde{e} \ne 0$.
Indeed, the  energy  of  the   nongauged  $Q$-ball  with  the  self-interaction
potential (\ref{II:5}) tends to infinity both in the  thin-wall regime $\tilde{
\omega} \rightarrow  \tilde{\omega}_{\min} = (1 - (3/16)\tilde{h}^{-1})^{1/2}$,
where $\tilde{E} \propto (\tilde{\omega}^{2} -\tilde{\omega}_{\min}^{2})^{-3}$,
and in the  thick-wall  regime $\tilde{\omega} \rightarrow 1$, where $\tilde{E}
\propto( 1 - \tilde{\omega} ^{2}) ^{-1/2}$.
At the same time, the  energy  of  any  gauged  $Q$-ball remains finite for all
$\tilde{\Omega}_{\infty}  \in  [\tilde{\Omega}_{\infty}^{\min}, 1]$,  where the
minimum possible value $\tilde{\Omega}_{\infty}^{\min}$ depends  on $\tilde{e}$
in accordance  with Fig.~\ref{fig2}.
We see again that for a  given  $\tilde{\Omega}_{\infty}$,  there exist the two
gauged $Q$-ball  solutions  whose  energies  can  differ  by  several orders of
magnitude.
In particular,  the  curves $\tilde{E}(\tilde{\Omega}_{\infty})$  intersect the
limiting line $\tilde{\Omega}_{\infty} = 1$ at two points.
It was found numerically that  for  the $Q$-ball solutions corresponding to the
upper intersection  points,  the  energy,  Noether  charge, and electric charge
increase indefinitely as $\tilde{e} \rightarrow 0$:
\begin{equation}
\tilde{E}\sim \bar{\varepsilon }\tilde{e}^{-3},\quad
\tilde{Q}_{N} \sim \bar{q}_{N}\tilde{e}^{-3},  \quad
\tilde{Q} \sim \bar{q}_{N}\tilde{e}^{-2},                          \label{IV:5}
\end{equation}
where $\bar{\varepsilon}$ and  $\bar{q}_{N}$ are dimensionless functions of the
self-interaction coupling constant $\tilde{h}$.
It can be said that for sufficiently small $\tilde{e}$, the gauged $Q$-ball that
corresponds to the upper intersection  point  passes  into  the quasi-thin-wall
regime described in Ref.~\cite{klee}.

The $\tilde{e}$ dependences in  Eq.~(\ref{IV:5}) can be explained qualitatively
as follows.
The basic property of the thin-wall regime is that the ansatz functions $\tilde{
f}(\tilde{r})$ and $\tilde{\Omega}(\tilde{r})$  are  approximately constant for
$\tilde{r} \in [0, \tilde{R})$, where $\tilde{R}$ can be arbitrarily  large  in
the  true  thin-wall  regime  (nongauged  case)  and  large  but finite in  the
quasi-thin-wall  regime (gauged case).
Eq.~(\ref{III:5a}) tells us that $\tilde{f}(\tilde{r})$  can be almost constant
on the interval $[0,\tilde{R})$ only if $(\tilde{\Omega}^{2}  -  1) \tilde{f} +
\tilde{f}^{3}/2 - \tilde{h} \tilde{f}^{5}/4 \approx 0$ there.
It follows that in the quasi-thin-wall regime
\begin{equation}
\tilde{f}_{0}\approx \tilde{h}^{-\frac{1}{2}}
\left[1 + \left(1 - 4 \tilde{h}
\left(1 - \tilde{\Omega}_{0}^{2}\right)
\right)^{\frac{1}{2}}\right]^{\frac{1}{2}},                       \label{IV:5a}
\end{equation}
where  $\tilde{\Omega}_{0}$   and   $\tilde{f}_{0}$   are  the  values  of  the
corresponding ansatz functions at $\tilde{r} = 0$.
However,   the    term    $\tilde{e}^{2}   \tilde{\Omega}   \tilde{f}^{2}$   in
Eq.~(\ref{III:5b}) does not allow the ansatz function  $\tilde{\Omega}(\tilde{r
})$  to  be  approximately  constant  on  the  interval $[0,\tilde{R})$ with an
arbitrary large  $\tilde{R}$; hence, there is no true thin-wall regime in model
(\ref{II:1})  and,  as  a  consequence, there  is no gauged $Q$-ball possessing
an arbitrarily large Noether (and, consequently, electric) charge.
From Eq.~(\ref{III:8b}), it follows that $\tilde{\Omega}\left(\tilde{r} \right)
\approx  \tilde{\Omega}_{0} + \tilde{e}^{2}\tilde{f}_{0}^{2} \tilde{\Omega}_{0}
\tilde{r}^{2}/6$.
We see  that  $\tilde{\Omega}\left( \tilde{r} \right)$  will  be  approximately
constant on the interval $[0,\tilde{R})$ until $\tilde{e} \tilde{f}_{0} \tilde{
R} \ll 1$.
It follows that
\begin{equation}
\tilde{R} \approx \varrho \tilde{f}_{0}^{-1} \tilde{e}^{-1},      \label{IV:5b}
\end{equation}
where the constant $\varrho \ll 1$.
Hence, the volume of the ball in which the ansatz  functions  $\tilde{\Omega}_{
0}$  and $\tilde{f}_{0}$ are approximately constant is proportional to $\tilde{
e}^{-3}$.
Next, Eqs.~(\ref{III:14}) and (\ref{III:15a}) tell us  that  the Noether charge
and energy densities are also  approximately constant inside the ball of radius
$\tilde{R} \propto \tilde{e}^{-1}$, and this gives rise to Eq.~(\ref{IV:5}).

In the  nongauged  case,  the  $Q$-ball  passes  into  the thick-wall regime as
$\tilde{\omega} \rightarrow 1$.
In this regime, the amplitude $\tilde{f}$ of the complex  scalar field tends to
zero as $(1-\tilde{\omega}^{2})^{1/2}$ and thus the nongauged $Q$-ball solution
spreads over the space.
It was  shown in  Ref.~\cite{paccetti} that for the potential (\ref{II:5}), the
energy and   Noether  charge  of  the  thick-wall nongauged $Q$-ball diverge as
$(1-\tilde{\omega}^{2})^{-1/2}$.
Conversely, it follows from  Fig.~\ref{fig5}  that  for  fixed $\tilde{e}$, the
energy of the  $Q$-ball  solution corresponding to the lower intersection point
remains finite at $\tilde{\omega} = 1$.
The amplitude $\tilde{f}$ also remains finite, and  thus there is no thick-wall
regime   when    the    gauge    coupling   constant   $\tilde{e}$   is   fixed
\cite{gulamov_2015}.
However, it was  found  numerically  that at the  lower intersection point, the
energy  and  Noether  charge  of  the  gauged  $Q$-ball  diverge  as $\tilde{e}
\rightarrow 0$:
\begin{equation}
\tilde{E}\approx \tilde{Q}_{N}\sim\bar{\bar{q}}_{N}\tilde{e}^{-1}, \label{IV:6}
\end{equation}
whereas the electric charge $\tilde{Q}=\tilde{e} \tilde{Q}_{N}$ remains finite.
This behaviour can be explained as follows.
In the unitary gauge, the  parameter  $\tilde{\Omega}_{\infty}$  plays the same
role as the phase frequency $\tilde{\omega}$ in the nongauged case.
From Fig.~\ref{fig3}, it  follows  that  at  $\tilde{\Omega}_{\infty} = 1$, the
difference  $\tilde{\mu}^{2} = 1 - \tilde{\Omega}_{0}^{2}$  tends  to  zero  as
$\tilde{e} \rightarrow 0$.
Next, it is shown  in  Appendix B  that  for  small  values  of  the  parameter
$\tilde{\mu}$, the Noether charge $\tilde{Q}_{N} \propto \tilde{\mu}^{-1}$.
However, the integral $\tilde{J}=\tilde{e}\int_{0}^{\infty}\tilde{r}\tilde{j}_{
0}\left(\tilde{r}\right)d\tilde{r}$ on the right-hand  side  of Eq.~(\ref{B:1})
is   proportional   to   $\tilde{e}^{2}$,   whereas   the   left-hand  side  of
Eq.~(\ref{B:1}) is proportional to $\tilde{\mu}^2/2$.
It follows that for small $\tilde{e}$, the parameter $\tilde{\mu}\propto\tilde{
e}$ and  thus $\tilde{E} \approx \tilde{Q}_{N} \propto \tilde{\mu}^{-1} \propto
\tilde{e}^{-1}$, in accordance with the numerical results.

\begin{figure}[t]
\includegraphics[width=0.5\textwidth]{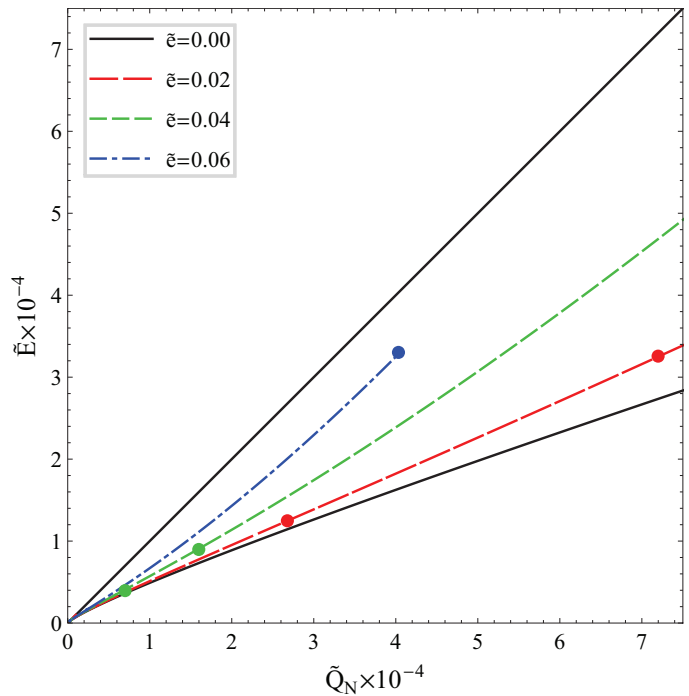}
\caption{\label{fig6}      Curves $\tilde{E}(\tilde{Q}_{N})$  for the first few
values of $\tilde{e}$ and $\tilde{h}=0.2$.   The inflection point (on the left)
and  the  point  of  contact  (on the right) are  shown  both  for $\tilde{e} =
0.02$ and $\tilde{e}  =  0.04$.  The terminal point is  shown  for $\tilde{e} =
0.06$. The curves correspond to the  unexcited  $Q$-ball  solution.}
\end{figure}

In Fig.~\ref{fig5}, all the curves (except  for the one  corresponding  to  the
nongauged case $\tilde{e} = 0$) have turning  points  at  which  the derivative
$d\tilde{E}/d\tilde{\Omega}_{\infty}$ becomes infinite.
Eq.~(\ref{III:3}) tells  us  that  $d\tilde{Q}_{N}/d\tilde{\Omega}_{\infty}$ is
also infinite at  the  turning  points, and thus the derivative $d\tilde{\Omega
}_{\infty}/d\tilde{Q}_{N}$ vanishes at these points.
This  means  that  the  second  derivative $d^{2}\tilde{E}/d\tilde{Q}_{N}^{2} =
d \tilde{\Omega}_{\infty}/d\tilde{Q}_{N}$ also vanishes at these turning points,
and thus the turning points on the  curves $\tilde{E}(\tilde{\Omega}_{\infty})$
correspond to the inflection points of  the  curves $\tilde{E}(\tilde{Q}_{N})$.
The curve $\tilde{E}(\tilde{Q}_{N})$ therefore has one inflection  point in the
gauged  case,  whereas  it  has  no  inflection  points  in the nongauged case.
These facts can be explained qualitatively as follows.
The derivative $d\tilde{E}/d\tilde{\Omega}_{\infty}$ determines the profitability
of  the  absorption  of  the  scalar  $\phi$-boson  into  the $Q$-ball from the
viewpoint of the energy balance.
Indeed, the  quantity  $m(1 - d\tilde{E}/d\tilde{Q}_{N})$  is the energy needed
to extract one scalar $\phi$-boson  from  the $Q$-ball with Noether charge $Q_{
N}$ and to transport the extracted $\phi$-boson to infinity.
From Eq.~(\ref{III:3}), it follows that the derivative $d\tilde{E}/d\tilde{Q}_{
N}$ decreases  monotonically  with   a  decrease  in $\tilde{\Omega}_{\infty}$.
It than follows from Fig.~\ref{fig5} that for the nongauged $Q$-ball solutions
lying on the left (thin-wall)  part  of  the  curve $\tilde{E}(\tilde{\Omega}_{
\infty})$, the  derivative  $d\tilde{E}/d\tilde{Q}_{N}$ decreases monotonically
with an  increase in $\tilde{Q}_{N}$ and thus the difference $(m+E(Q_{N} - 1))-
E(Q_{N})\approx m(1 - d\tilde{E}/d\tilde{Q}_{N})$  increases monotonically with
an increase in $\tilde{Q}_{N}$.
We conclude that in the nongauged case $\tilde{e}  =  0$, the absorption of the
$\phi$-boson becomes more energetically profitable  with   an  increase  in the
$Q$-ball's Noether charge towards the thin-wall regime.
The situation changes drastically  for gauged $Q$-balls.
In this case, the role of  the  electrostatic  Coulomb repulsion increases with
an increase in the Noether  (and consequently the electric) charge.
The long-range Coulomb repulsion means that the absorption  of the $\phi$-boson
into  a  gauged  $Q$-ball  with  Noether  charge  $Q_{N}$  is  less  profitable
energetically  in  comparison  with  the  absorption  of  the $\phi$-boson into
a nongauged $Q$-ball with the same $Q_{N}$.
Hence, for the same $Q_{N}$, the difference $(m+E(Q_{N} - 1))- E(Q_{N}) \approx
m(1-d\tilde{E}/d\tilde{Q}_{N})$ is smaller for the gauged $Q$-ball than for the
nongauged  one,  and  thus $d\tilde{E}_{\text{g}}/d\tilde{Q}_{N} > d\tilde{E}_{
\text{ng}}/d\tilde{Q}_{N}$, where $\tilde{E}_{\text{g}}$ and $\tilde{E}_{\text{
ng}}$ are the energies of  the  gauged  and  nongauged $Q$-balls, respectively.
The difference  $d\tilde{E}_{\text{g}}/d\tilde{Q}_{N} - d\tilde{E}_{\text{ng}}/
d\tilde{Q}_{N}$ increases  with  an   increase  in  $\tilde{Q}_{N}$, leading to
the inflection  points on the curves $\tilde{E}(\tilde{Q}_{N})$  in  the gauged
case (see Fig.~\ref{fig6})  and, as  a  consequence,  to  the turning points on
the curves $\tilde{E}(\tilde{\Omega}_{\infty})$ in Fig.~\ref{fig5}.
In Appendix C, the existence of  the  inflection  point on the curve $\tilde{E}
(\tilde{Q}_{N})$ is  explained  analytically  within  a  certain approximation.

\begin{figure}[t]
\includegraphics[width=0.5\textwidth]{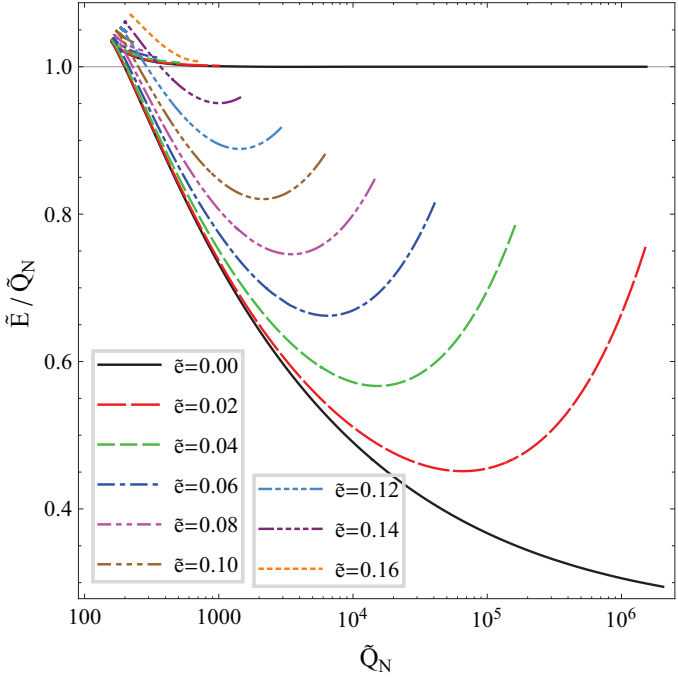}
\caption{\label{fig7}    Dependence of  the  ratio $\tilde{E}/\tilde{Q}_{N}$ on
the  Noether  charge  $\tilde{Q}_{N}$  for  different  values  of   $\tilde{e}$
and  $\tilde{h} = 0.2$.   The  curves  correspond  to  the  unexcited  $Q$-ball
solution.}
\end{figure}

Figure~\ref{fig6} presents  the curves $\tilde{E}(\tilde{Q}_{N})$ for the first
few   values   of   the   gauge   coupling   constant   $\tilde{e}$   and   the
self-interaction coupling  constant $\tilde{h} = 0.2$.
For $\tilde{e} = 0.02$ and $\tilde{e} = 0.04$, two characteristic points of the
curve  $\tilde{E}(\tilde{Q}_{N})$   are  shown:  the  inflection  point,  where
$d^{2}\tilde{E}/d\tilde{Q}_{N}^{2}  =  0$,  and  the  point  of  contact, where
$d\tilde{E}/d\tilde{Q}_{N}  =  \tilde{E}/\tilde{Q}_{N}$.
The inflection points lie to the left of  the corresponding points of  contact.
For  $\tilde{e} = 0.06$,  the  terminal  point  corresponding  to  the  maximum
allowable Noether charge and energy is shown,  whereas the inflection point and
the point of contact are not shown due to their indistinguishability.
At  the  terminal  point,  the  derivative $d\tilde{E}/d\tilde{Q}_{N}  = 1$, in
accordance with Eq.~(\ref{III:3}) and Fig.~\ref{fig5}.
From Fig.~\ref{fig6}, it  follows  that the $Q$-ball's energy increases rapidly
both with  an  increase in $\tilde{e}$  for  fixed  $\tilde{Q}_{N}$ and with an
increase in $\tilde{Q}_{N}$ for fixed $\tilde{e}$.

Figure~\ref{fig7} presents the dependence of the ratio $\tilde{E}/\tilde{Q}_{N}$
on the  Noether  charge  $\tilde{Q}_{N}$  for  different  values  of  the gauge
coupling constant $\tilde{e}$.
The curves in  Fig.~\ref{fig7} correspond  to the unexcited $Q$-ball solutions.
The corresponding  curves  for  the  radially  excited  $Q$-ball  solutions are
similar to those in Fig.~\ref{fig7}, and thus are not shown.
As in  the  previous  figures,  the  curve  for  the nongauged $Q$-ball differs
sharply from those for the gauged $Q$-balls.
Firstly,  we  see  that  in  accordance  with  Fig.~\ref{fig5}, the  energy and
Noether  charge  of   the   nongauged   $Q$-ball  increase  indefinitely,  both
in the thin-wall ($\tilde{E}/\tilde{Q}_{N} \rightarrow  \tilde{\omega}_{\min}$)
and in the thick-wall ($\tilde{E}/\tilde{Q}_{N}\rightarrow 1$) regimes, whereas
there are  upper  bounds  on  the  energy  and  Noether  charge  of  the gauged
$Q$-balls.
Next,  in  Fig.~\ref{fig7},  all  the  curves  (except those that correspond to
$\tilde{e} = 0.14$ and $\tilde{e} = 0.16$)  have  cuspidal points; these points
correspond to the minima of the  curves $\tilde{E}(\tilde{\Omega}_{\infty})$ in
Fig.~\ref{fig5}.
Note that due to  Eq.~(\ref{III:3}),  the  position of the minimum in the curve
$\tilde{E}(\tilde{\Omega}_{\infty})$  coincides with that  in the corresponding
curve  $\tilde{Q}_{N}(\tilde{\Omega}_{\infty})$,  giving  rise  to the cusps in
Fig.~\ref{fig7}.
In  Fig.~\ref{fig5},  the   curves   $\tilde{E}(\tilde{\Omega}_{\infty})$  with
$\tilde{e}=0.14$ and $\tilde{e} =0.16$ have no minima, resulting in the absence
of cusps for the corresponding curves in Fig.~\ref{fig7}.
We see that for all of the curves,  the ratio $\tilde{E}/\tilde{Q}_{N}$ reaches
a maximum value at the point  (cuspidal or otherwise) with the minimum possible
Noether charge $\tilde{Q}_{N}$.

In Fig.~\ref{fig7}, all the curves  corresponding  to  gauged $Q$-balls (except
the one corresponding to $\tilde{e} = 0.16$)  have  global  minima, whereas the
curve that corresponds to the nongauged $Q$-ball has no minimum.
It can  easily be shown that at the minimum point, $d\tilde{E}/d\tilde{Q}_{N} =
\tilde{E}/\tilde{Q}_{N}$ and $d^{2}(\tilde{E}/\tilde{Q}_{N})/d\tilde{Q}_{N}^{2}
= (d^{2}\tilde{E}/d\tilde{Q}_{N}^{2})/\tilde{Q}_{N}$.
From Fig.~\ref{fig7}, it follows  that the derivative $d^{2}(\tilde{E}/\tilde{Q
}_{N})/d\tilde{Q}_{N}^{2}$  is  positive  at  the  minimum points, and thus the
derivative $d^{2}\tilde{E}/d\tilde{Q}_{N}^{2}$ is also positive there.
Hence, the curve $\tilde{E}(\tilde{Q}_{N})$ has a convex downwards shape at the
minimum point  of  $\tilde{E}(\tilde{Q}_{N})/\tilde{Q}_{N}$; this is consistent
with  the  existence of  inflection  points  for   the   curves  with  non-zero
$\tilde{e}$ in Fig.~\ref{fig6}.
It also follows from Fig.~\ref{fig6} that the curve for zero $\tilde{e}$ has no
inflection  point, due to the absence of a minimum point for the curve for zero
$\tilde{e}$ in Fig.~\ref{fig7}.

At the minimum point of the curve $\tilde{E}(\tilde{Q}_{N})/\tilde{Q}_{N}$, the
derivative $d\tilde{E}/d\tilde{Q}_{N} =\tilde{E}/\tilde{Q}_{N}$, and this point
therefore corresponds to  the  point  of  contact between the curve $\tilde{E}(
\tilde{Q}_{N})$ and the straight line passing through the origin of coordinates.
From Fig.~\ref{fig6}, it follows that the  inflection  point  lies  to the left
of the point of contact.
Hence, it is  possible  that  the  curve  $\tilde{E}(\tilde{Q}_{N})$ terminates
after the inflection point but has  no  point of contact with the straight line
passing through the origin of coordinates.
In Fig.~\ref{fig7}, this situation is seen for the curve with $\tilde{e}=0.16$,
which has no minimum.

\begin{figure}[t]
\includegraphics[width=0.5\textwidth]{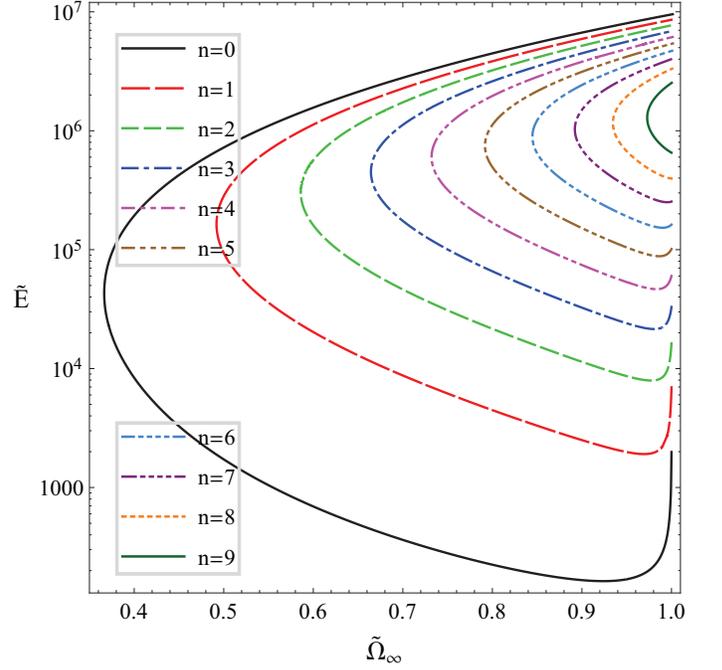}
\caption{\label{fig8}     Curves  $\tilde{E}(\tilde{\Omega}_{\infty})$  for the
unexcited  and  first  nine  radially  excited  $Q$-ball solutions.  The curves
correspond to parameters $\tilde{e} = 0.01$ and $\tilde{h} = 0.2$.}
\end{figure}

In Fig.~\ref{fig7}, the  cuspidal points divide the curves into upper and lower
branches, and the curves without cuspidal points (corresponding to $\tilde{e} =
0.14$ and $\tilde{e}=0.16$) have only one branch.
We  see  that  for   non-zero  $\tilde{e}$,  both  the  lower  branches  of the
double-branch  curves  and  the single-branch curves terminate at the rightmost
points.
These points  correspond  to  the  $Q$-ball  solutions  with maximum energy and
Noether charge for a given $\tilde{e}$.
From Eq.~(\ref{III:3})  and  Figs.~\ref{fig3}  and  \ref{fig5}, it follows that
for these $Q$-ball  solutions,  the  derivative $d\tilde{E}/d\tilde{Q}_{N} = 1$
(and consequently $dE/dQ_{N} = m$).
Similarly, the  derivative  $d\tilde{E}/d\tilde{Q}_{N}$  is also equal to unity
at  the rightmost points of  the  upper  branches  of  the  curves  and  at the
leftmost points of the single-branch curves.
We see that the gauged  $Q$-balls  cease  to  exist  at  the  points, where the
derivative  $dE/dQ_{N}$  is  equal  to  the  mass  of  the  scalar $\phi$-boson
\cite{klee}.
Of course, the reason for this lies  in  Eqs.~(\ref{III:3}) and (\ref{III:11}).
The former relation  tells us that the  parameter  $\Omega_{\infty} = m$ at the
points  where the derivative $dE/dQ_{N} = m$, while  the  latter means that the
amplitude of the complex scalar field becomes oscillating when $\Omega_{\infty}
> m$.
This implies that the Noether  charge  and  energy become infinite as $\Omega_{
\infty} > m$, and thus the $Q$-ball solution does not exist in this case.

It  follows  from   Fig.~\ref{fig7} that  at  fixed $\tilde{Q}_{N}$,  the ratio
$\tilde{E}/\tilde{Q}_{N}$  increases   with  an  increase  in  $\tilde{e}$  for
$Q$-ball solutions  both on upper and lower  branches  of the curves $\tilde{E}
(\tilde{Q}_{N})/\tilde{Q}_{N}$.
This is apparently  related  to  the increase in the role of long-range Coulomb
repulsion. 
We also see that  the  ratio  $\tilde{E}/\tilde{Q}_{N}$  is less than unity for
$Q$-balls with the  maximum  possible  energy  and  Noether  charge (except for
$\tilde{e} = 0.16$).
It was found numerically that for these $Q$-balls, the ratio $\tilde{E}/\tilde{
Q}_{N}$ tends to a constant as  $\tilde{e}  \rightarrow 0$, and the constant is
approximately  equal to $0.73$.
At the same time, for the nongauged $Q$-ball, the ratio $\tilde{E}/\tilde{Q}_{N
}$ tends to $\tilde{\omega}_{\min} = (1 - (3/16)\tilde{h}^{-1})^{1/2}  =  0.25$
when $\tilde{E}$ and $\tilde{Q}_{N}$ tend to infinity  in the thin-wall regime.
It follows that for gauged $Q$-balls, the ratio  $\tilde{E}/\tilde{Q}_{N}$ does
not tend to the nongauged value of $0.25$ as $\tilde{e} \rightarrow 0$.
Hence, in the limit of vanishing $\tilde{e}$, the gauge field does not decouple
from  the $Q$-ball with the maximum possible  energy   and   Noether  charge in
consistency with the conclusion obtained in analysis of Fig.~\ref{fig3}.
Note that for the gauged $Q$-ball  with  an  arbitrarily  small $\tilde{e}$ and
maximum   possible  $\tilde{Q}_{N}$,  the  ratio  $\tilde{E}/\tilde{Q}_{N}$  is
distinctly greater  than  that  for  the  nongauged  $Q$-ball  in the thin-wall
regime. 
Of course, this difference is  due  to  the  contribution  of the electrostatic
energy (\ref{III:22a}) to the total energy of gauged $Q$-ball.

Figure~\ref{fig8} shows the curves $\tilde{E}(\tilde{\Omega}_{\infty})$ for the
unexcited  and  first  nine  radially  excited  $Q$-ball solutions.
As in Fig.~\ref{fig7}, all the curves in Fig.~\ref{fig8} intersect the limiting
line  $\tilde{\Omega}_{\infty} = 1$  at two points and possess  turning points.
We see  that in  the  same  way  as  in  Fig.~\ref{fig2}, the $\tilde{\Omega}_{
\infty}$ coordinate  of  the turning point increases  monotonically towards the
limiting value $\tilde{\Omega}_{\infty} = 1$ with an increase in $n$.
We also see that the maximum possible energy of radially excited gauged $Q$-ball
decreases monotonically with an increase in $n$.
Note that  such  behavior is similar  to  that  shown in Fig.~\ref{fig5}, where
the  maximum  possible  energy  of  the  unexcited  gauged  $Q$-ball  decreases
monotonically with an  increase  in  the  gauge  coupling constant $\tilde{e}$.

\begin{figure}[t]
\includegraphics[width=0.5\textwidth]{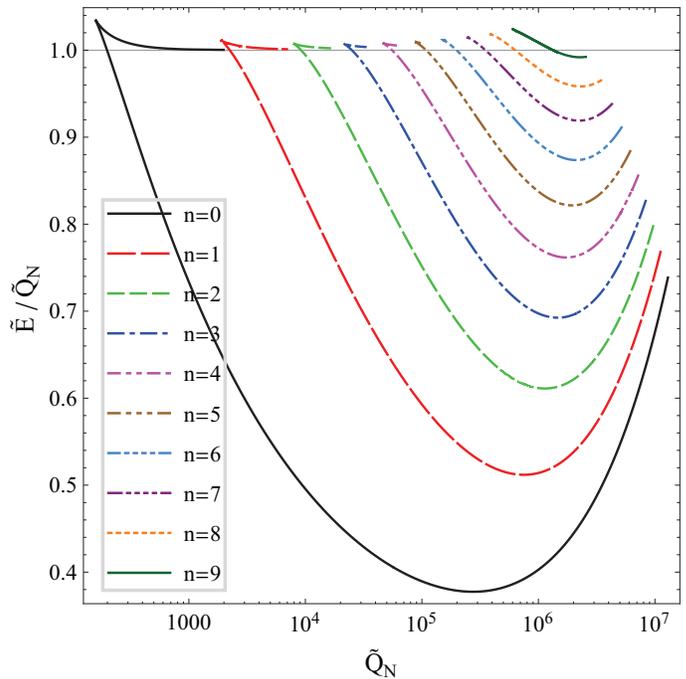}
\caption{\label{fig9}      Dependence of the ratio $\tilde{E}/\tilde{Q}_{N}$ on
the Noether charge $\tilde{Q}_{N}$  for  the unexcited and  first nine radially
excited $Q$-ball solutions.    The curves correspond to parameters $\tilde{e} =
0.01$ and $\tilde{h} = 0.2$.}
\end{figure}

\begin{figure}[t]
\includegraphics[width=0.5\textwidth]{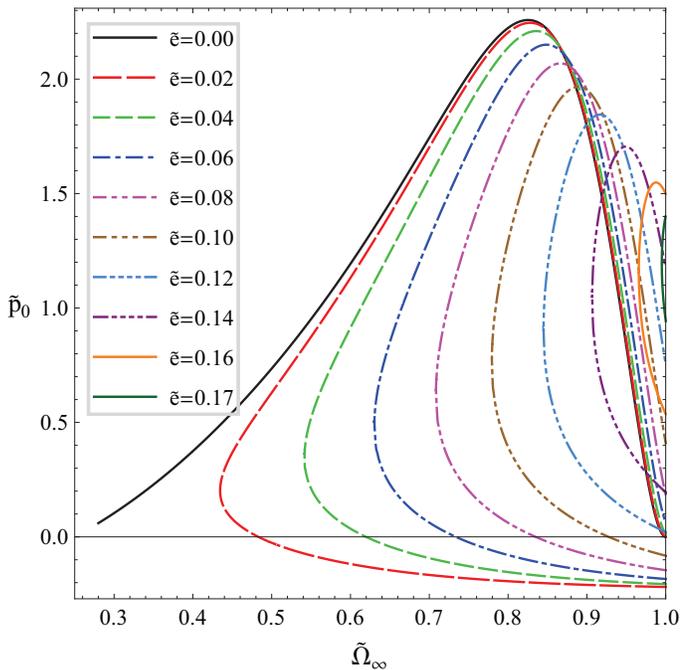}
\caption{\label{fig10}     Curves $\tilde{p}_{0}(\tilde{\Omega}_{\infty})$ for
different values of $\tilde{e}$ and $\tilde{h} = 0.2$.   The curves correspond
to the unexcited $Q$-ball solution.}
\end{figure}

Next, we turn  to  Fig.~\ref{fig9},  in   which  the  dependence  of  the ratio
$\tilde{E}/\tilde{Q}_{N}$ on the  Noether  charge  $\tilde{Q}_{N}$ is shown for
the unexcited and first nine radially excited $Q$-ball solutions.
We see that the  individual  behaviour  of  the  curves  in  Fig.~\ref{fig9} is
similar to that in Fig.~\ref{fig7}.
However, the relative  positions of the curves in Fig.~\ref{fig9} are different
from those in Fig.~\ref{fig10}, since the cusps  (or the leftmost points for $n
= 8,\,9$) in Fig.~\ref{fig9} are separated by larger intervals of $\tilde{Q}_{N
}$ than those in Fig.~\ref{fig7}.
All the  curves  $\tilde{E}( \tilde{Q}_{N} )/\tilde{Q}_{N}$  in Fig.~\ref{fig9}
(except the one corresponding to $n = 9$) have minimum points.
With the existence of the turning points in  Fig.~\ref{fig8}, this implies that
the  all  corresponding  curves  $\tilde{E}(\tilde{Q}_{N})$   have   inflection
points at  which  $d^{2}\tilde{E}/d\tilde{Q}_{N}^{2} = 0$  and (except for $n =
9$)  points   of  contact  at  which  $d\tilde{E}/d\tilde{Q}_{N}  =  \tilde{E}/
\tilde{Q}_{N}$.
All the curves in Fig.~\ref{fig9} possess  two  terminal  points  corresponding
to the two points of intersection with the  line $\tilde{\Omega}_{\infty}=1$ in
Fig.~\ref{fig8}.
This implies that at the terminal points,  the derivative $d\tilde{E}/ d\tilde{
Q}_{N} = 1$.
Note that in Fig.~\ref{fig9}, the ratio $\tilde{E}/\tilde{Q}_{N}$  is less than
unity at all of  the  terminal  points at the right; for a given $n$, the right
terminal point corresponds to the  $Q$-ball solution with the maximum allowable
energy and Noether charge.
Finally, it follows from Fig.~\ref{fig9}   that  for fixed $\tilde{Q}_{N}$, the
energy of the $Q$-ball solution increases with an increase in $n$.

Let us discuss the stability of radially excited gauged $Q$-balls.
Figure~\ref{fig9}  tells  us  that  for  the  $n$-th  radially  excited  gauged
$Q$-ball, there  are  at least the $n$ gauged $Q$-balls having the same Noether
charge but lower energy.
It follows  that  all  radially  excited  gauged  $Q$-balls  are  unstable with
respect to the transition into less excited ones.
The energy released in  this  transition is carried away by the electromagnetic
radiation and the scalar $\phi$-bosons.
Moreover, from Fig.~\ref{fig9} it  follows  that  for all $n$, there are gauged
$Q$-ball solutions (those for which the ratio $\tilde{E}/\tilde{Q}_{N}>1$) that
are unstable against decay into the massive scalar $\phi$-bosons.
Thus, all the radially excited gauged $Q$-balls are unstable.
The instability, however, can be either classical  (the presence of one or more
unstable  modes  in   the   functional   neighborhood  of  gauged $Q$-ball)  or
quantum-mechanical (the possibility of quantum tunneling).
Note that  the  issue  of  classical  stability  of  gauged $Q$-balls is rather
complicated.
Indeed,  unlike  the  nongauged   case,   there  is  no  criterion of classical
stability for gauged $Q$-balls \cite{panin_2017}.
Note,  however,  that  in   Fig.~\ref{fig9},  there   are  cusps  on the curves
corresponding to $n = 0, \ldots, 6$.
It was shown  in \cite{fried_lee_77, lee_pang_89} that the appearance of a cusp
indicates the onset of a new mode of instability.
Hence, the curves with cusp have areas of classical instability.
At the same time, the presence of cusp may mean  the addition of a new unstable
mode to already existing ones.
In this case, all  the  radially  excited gauged $Q$-balls would be classically
instable as it is in the model \cite{fried}.

\begin{figure}[t]
\includegraphics[width=0.5\textwidth]{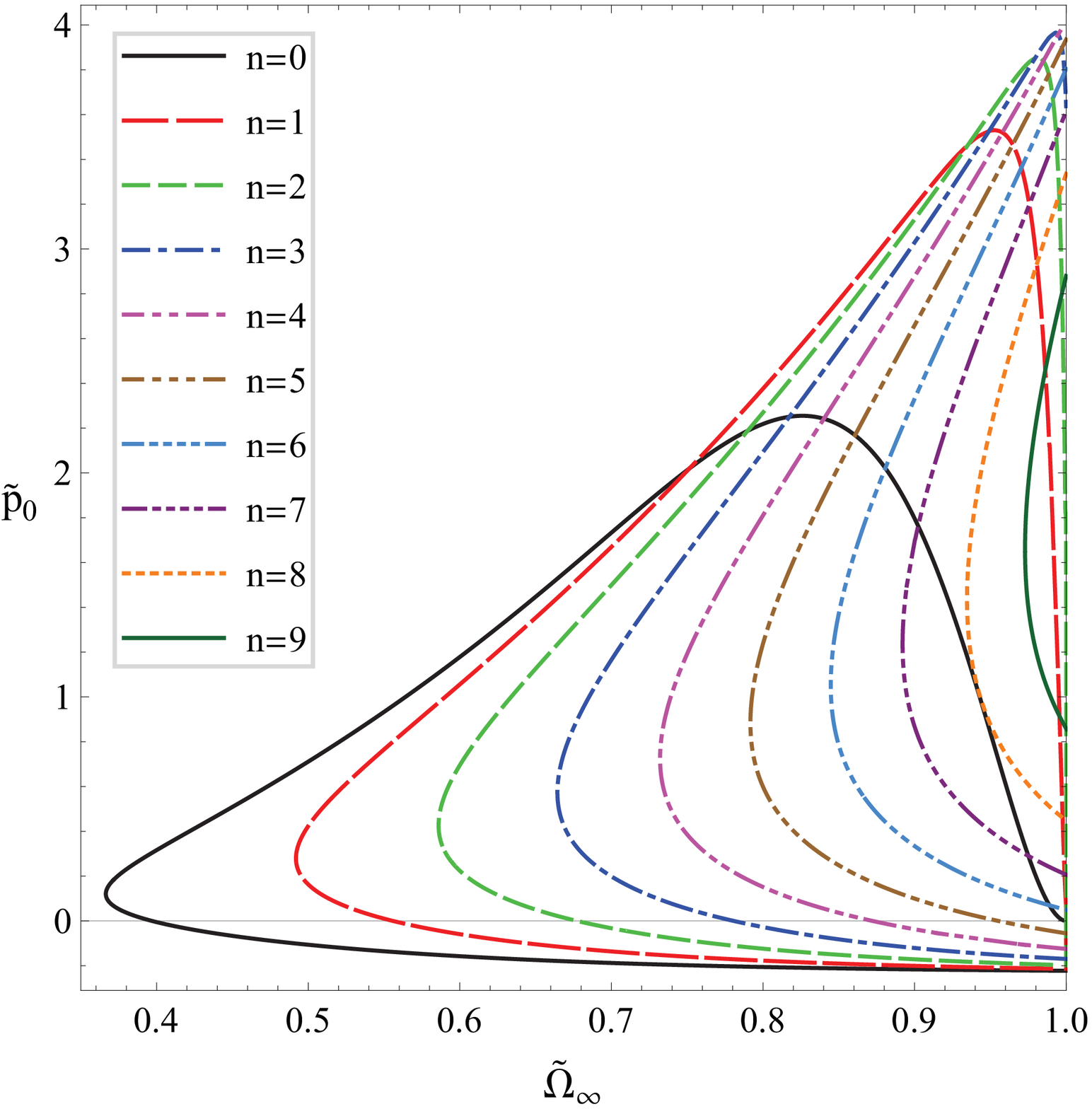}
\caption{\label{fig11}     Curves  $\tilde{p}_{0}(\tilde{\Omega}_{\infty})$ for
the  unexcited  and  first  nine   radially   excited  $Q$-ball solutions.  The
curves correspond to  parameters $\tilde{e}=0.01$ and $\tilde{h} = 0.2$.}
\end{figure}

The boundary conditions (\ref{III:6}) and  Eq.~(\ref{III:16b}) tell us that the
pressure $p_{0}$ at the center of the $Q$-ball solution can be written as
\begin{equation}
p_{0}\equiv m^{4}g^{-1}\tilde{p}_{0}  =  m^{4}g^{-1}\left[\frac{1}{2}
\tilde{\Omega }_{0}^{2}\tilde{f}_{0}^{2}-V(\tilde{f}_{0}) \right], \label{IV:7}
\end{equation}
where $\tilde{\Omega}_{0}=\tilde{\Omega }(0)$ and $\tilde{f}_{0}=\tilde{f}(0)$.
In Eq.~(\ref{IV:7}), the  expression  in  square  brackets is the dimensionless
version of the effective potential (\ref{B:5}): $\tilde{U}_{\text{eff}}(\tilde{
f}) = \tilde{\Omega}^{2}\tilde{f}^{2}/2 - V(\tilde{f})$.
In  the  nongauged  case,  the   ansatz   function  $\tilde{\Omega}(\tilde{r})$
becomes the constant phase frequency $\tilde{\omega}$.
It is well known  that  in  this  case, the value of the effective potential is
always positive at the center of the $Q$-ball solution;  hence, the pressure at
the center of the nongauged $Q$-ball solution is also always positive.
The situation  is  different  in  the  gauged  case,  however, as the effective
potential $\tilde{U}_{\text{eff}}(\tilde{f})$ may be negative at $\tilde{r}=0$,
resulting  in    negative   pressure   at  the  center  of the gauged  $Q$-ball
\cite{gulamov_2015}.
Figures~\ref{fig10}  and  \ref{fig11}  show  the  dependences  of  the  central
pressure $\tilde{p}_{0}$ on the parameter $\tilde{\Omega}_{\infty}$.
The former  shows  the  curves $\tilde{p}_{0}(\tilde{\Omega}_{\infty})$ for the
unexcited $Q$-ball  solutions  corresponding  to  different  values of $\tilde{
e}$, while  the  latter  shows  the  curves  for  the  unexcited and first nine
radially excited  $Q$-ball solutions for fixed $\tilde{e}=0.01$.
All the curves in  Figs.~\ref{fig10}  and \ref{fig11}  have  similar behaviour.
They intersect the limiting line $\tilde{\Omega}_{\infty} = 1$ at the upper and
lower points, and possess turning points.
The only  exception  is  the  curve $\tilde{p}_{0}(\tilde{\Omega}_{\infty})$ in
Fig.~\ref{fig10}, which corresponds to the nongauged case $\tilde{e} = 0$; this
curve has only one intersection point, and does not have a turning point.
Note that the positions of  the turning points in Figs.~\ref{fig3}, \ref{fig5},
and \ref{fig10}  coincide,  as  do  the  positions  of  the  turning  points in
Figs.~\ref{fig4}, \ref{fig8}, and \ref{fig11}.
However, the main feature of Figs.~\ref{fig10} and  \ref{fig11} is the presence
of rather broad  intervals  of  $\tilde{\Omega}_{\infty}$  in which the central
pressure of the $Q$-ball solution is negative.
This behavior is  characteristic  for  curves corresponding to $\tilde{e}=0.02,
\, 0.04,\, 0.06,\, 0.08$,  and  $0.1$  in  Fig.~\ref{fig10}, and for the curves
corresponding to $n = 0, \ldots, 5$  in Fig.~\ref{fig11}.
Note  that   a    negative   central   pressure   is   characteristic  for  the
quasi-thin-wall regime, i.e., for  gauged  $Q$-ball  solutions possessing large
energies and Noether charges.
This  is  because  the  central  effective  potential  $\tilde{U}_{\text{eff}}$
(and  consequently  the  central  pressure) of   large  nongauged  $Q$-balls is
only slightly higher than zero, and  thus  may  become negative when we turn on
the electromagnetic interaction.
Finally, of all the curves  in  Figs.~\ref{fig10} and \ref{fig11}, only the one
corresponding to the  nongauged  $Q$-ball  in Fig.~\ref{fig10} tends to zero at
both (left and right) terminal points.
This is because the effective potential of the nongauged $Q$-ball tends to zero
in both  the  thin-wall  ($\tilde{\Omega}_{\infty} \rightarrow [1-(3/16)\tilde{
h}^{-1}]^{1/2}$)  and   thick-wall  ($\tilde{\Omega}_{\infty}  \rightarrow  1$)
regimes.

\begin{figure}[tbp]
\includegraphics[width=0.5\textwidth]{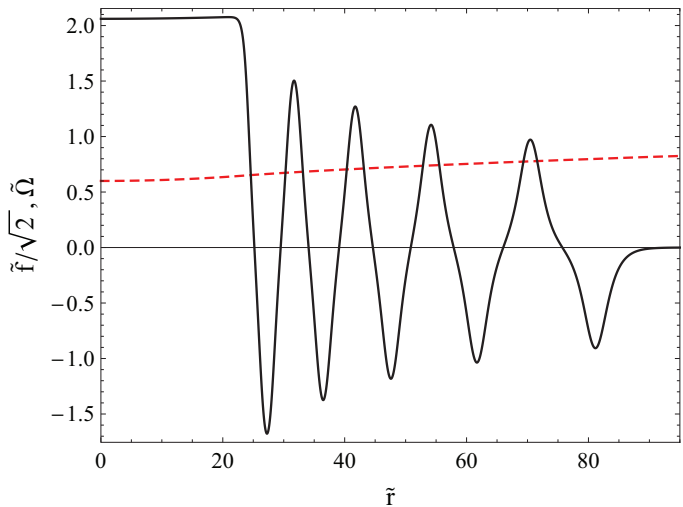}
\caption{\label{fig12a}       Profile functions $\tilde{f}(\tilde{r})/\sqrt{2}$
(solid  line)  and  $\tilde{\Omega}(\tilde{r})$ (dashed  line)  for  the  ninth
($n=9$)  radially  excited  $Q$-ball solution.      The solution corresponds to
parameters $\tilde{e} = 0.01$, $\tilde{h} = 0.2$,  and $\tilde{\Omega}_{\infty}
= 0.9817$.}
\end{figure}

A detailed description of the forms of the radially excited  nongauged $Q$-ball
solutions is given in Ref.~\cite{mai_2012}.
We have found that the forms of the  ansatz  function  $f(r)=mg^{-1/2}\tilde{f}
(\widetilde{r})$,  the   energy  density  $\mathcal{E}(r) = m^{4}g^{-1} \tilde{
\mathcal{E}}(\tilde{r})$, the  Noether charge density $j_{N}^{0}(r)=m^{3}g^{-1}
\tilde{j}_{N}^{0}(\tilde{r})$,  and  the  pressure $p(r) = m^{4}g^{-1}\tilde{p}
(\tilde{r})$ for the  gauged  case are similar to those for the nongauged case.
We give  as   an   example   only   the   excited  $Q$-ball  solution  shown in
Fig.~\ref{fig12a}, corresponding to the parameters $\tilde{e}=0.01$, $\tilde{h}
=0.2$, $n=9$, and  $\tilde{\Omega}_{\infty}=0.9817$,  as it demonstrates all of
the characteristic properties  of  radially excited gauged  $Q$-ball solutions.
Note that the solution  in  Fig.~\ref{fig12a}  is the maximum possible radially
excited solution for a given $\tilde{e}$ and $\tilde{h}$.
We can see that the  ansatz  function $\tilde{f}(\tilde{r})$ has the nine nodes
($n = 9$) that separate the nine alternating peaks.
At the same time,  the  ansatz  function  $\tilde{\Omega}(\tilde{r})$ increases
monotonically   on    the   interval   $[0,  \infty)$,   in   accordance   with
Eq.~(\ref{III:7a}).

\begin{figure}[tbp]
\includegraphics[width=0.5\textwidth]{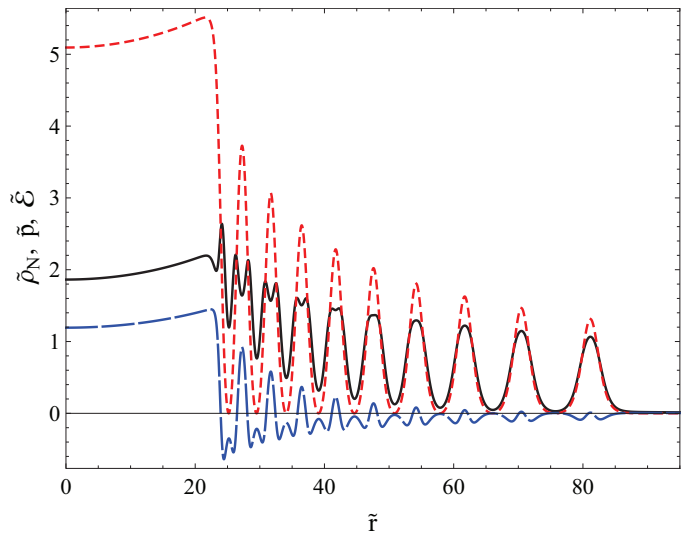}
\caption{\label{fig12b}       Energy  density  $\tilde{\mathcal{E}}(\tilde{r})$
(solid line),  Noether  charge  density  $\tilde{\rho}(\tilde{r}) = \tilde{j}_{
N}^{0}(\tilde{r})$  (short-dashed line),  and   pressure $\tilde{p}(\tilde{r})$
(long-dashed line) corresponding to the $Q$-ball solution in Fig.~\ref{fig12a}.}
\end{figure}

It was found  numerically that for $n$ close to the maximum possible value, the
absolute  values   of   $\tilde{f}$   at  the  peak  positions  $\tilde{r}_{i}$
approximately satisfy the relation
\begin{equation}
|\tilde{f}(\tilde{r}_{i})|\tilde{r}_{i}^{\frac{1}{2}}
\approx \text{const}.                                              \label{IV:8}
\end{equation}
This can be explained as follows.
It is shown in Appendix B (see Eqs.~(\ref{B:4}) and (\ref{B:5})) that the system
of  differential  equations  (\ref{III:5a})  and  (\ref{III:5b})  describes the
two-dimensional motion of a particle  with  unit mass in the plane $(\tilde{f},
\tilde{\Omega})$.
The particle moves in the force field of  the  effective potential $\tilde{U}_{
\text{eff}}(\tilde{f})  =  \tilde{\Omega}^{2}\tilde{f}^{2}/2  -  V(\tilde{f})$,
and   experiences   the   action   of   the   nonconservative   force  $\tilde{
\mathbf{F}}_{\text{nc}} = (0,  (1 + \tilde{e}^{2})\tilde{\Omega}\tilde{f}^{2})$
and  the  friction  force  $\tilde{\mathbf{F}}_{\text{f}} = (- 2 \tilde{r}^{-1}
\tilde{f}^{\prime }, -2 \tilde{r}^{-1} \tilde{\Omega}^{\prime})$.
Next, the  amplitudes  of  the  peaks  in  Fig.~\ref{fig12a}  are determined by
the effective potential $\tilde{U}_{\text{eff}}(\tilde{f})$.
In the wide neighbourhood  of  the point $(\tilde{f}, \tilde{\Omega}) = (0,1)$,
the contour  lines   corresponding  to  zero  levels  of  $U_{\text{eff}}$  and
$\partial_{\tilde{f}}U_{\text{eff}}$  are  well  described  by  the approximate
expressions $\tilde{f}\approx 2 (1-\tilde{\Omega})^{1/2}$ and $\tilde{f}\approx
2\sqrt{2}(1-\tilde{\Omega})^{1/2}$, respectively.
During the oscillations,  the  particle  must  intersect with the zero level of
$\partial_{\tilde{f}}U_{\text{eff}}$; otherwise, there  is  no  restoring force
and oscillations are impossible.
At the  same  time,  the  ``coordinate'' $\tilde{f}$  of  the  particle  cannot
significantly exceed the zero level of $U_{\text{eff}}$, otherwise oscillations
are also impossible.
We see that in the  neighborhood  of  the  point $(\tilde{f}, \tilde{\Omega}) =
(0, 1)$,  the   amplitudes   of  $\tilde{f}$  oscillations   are  approximately
proportional to $(1-\tilde{\Omega})^{1/2}$.
Eq.~(\ref{III:10}) also tells  us that in the neighborhood of $\tilde{\Omega} =
1$, the  difference $1 - \tilde{\Omega} \propto \tilde{r}^{-1}$.
A combination of the last two expressions results in Eq.~(\ref{IV:8}).

Figure~\ref{fig12b} shows  the  distributions  of  the  energy density $\tilde{
\mathcal{E}}(\tilde{r})$, the Noether charge density $\tilde{j}_{N}^{0}(\tilde{
r})$,  and the pressure $\tilde{p}( \tilde{r} )$ corresponding to the  $Q$-ball
solution in Fig.~\ref{fig12a}.
All three distributions consist of  a  central region, where $\tilde{\mathcal{E
}}$,  $\tilde{j}_{N}^{0}$,  and  $\tilde{p}$   weakly  depend  on  $\tilde{r}$,
followed by a region in  which  $\tilde{\mathcal{E}}$, $\tilde{j}_{N}^{0}$, and
$\tilde{p}$ oscillate.
However, the  oscillation  patterns of $\tilde{\mathcal{E}}$, $\tilde{j}_{N}^{0
}$, and $\tilde{p}$ are different.
The Noether charge  density  $\tilde{j}_{N}^{0}(\tilde{r})$  has  the  simplest
oscillation pattern, consisting of a sequence of peaks separated by zero minima.
The positions of the zeros  in  $\tilde{j}_{N}^{0}$  coincide with those of the
ansatz function $\tilde{f}$,  whereas  the  maxima  in  $\tilde{j}_{N}^{0}$ are
slightly different from those of $\tilde{f}$ due to the inconstancy of $\tilde{
\Omega}$ in Eq.~(\ref{III:14}).
From Eq.~(\ref{IV:8}), it follows that the height of the $i$-th  peak of $\tilde{
j}_{N}^{0}$  is  approximately  inversely  proportional to its  radial position
$\tilde{r}_{i}$, whereas  from  Fig.~\ref{fig12b}  it follows that the width of
the $i$-th peak increases  moderately  with  an  increase in $i$ as well as the
distance $\tilde{r}_{i}  -  \tilde{r}_{i  -  1}$  between  neighboring  peaks.

The pattern of oscillation  of  the  energy  density $\tilde{\mathcal{E}}$ than
that of the  Noether charge density $\tilde{j}_{N}^{0}$.
As in the previous case, it is a sequence of peaks separated by minima, but the
peaks have a complicated structure and the minima are non-zero.
For $i \le 4$, there  are  small,  sharp  peaks  against  a background of wider
basic peaks.
This complicated pattern  of  peaks  is  due  to  the  interference between the
gradient (\ref{III:22b}), kinetic (\ref{III:22c}), and potential (\ref{III:22d})
parts of the energy  density, where  the  small,  sharp  peaks  arise  from the
contribution of the gradient part.
With an  increase in  $i$,  the contribution of the gradient part decreases due
to the decrease  in  the  derivative $\tilde{f}'(\tilde{r})$; this  results  in
the disappearance of the small, sharp peaks, starting with $i = 5$.
The values of $\tilde{\mathcal{E}}$ at the  minimum  points located between the
wide basic peaks are non-zero and  result  from  the  contributions
of the electric (\ref{III:22a})   and   gradient  (\ref{III:22b})  parts, where
the contribution of the latter predominates.
These values therefore allow  us  to estimate the relative contributions of the
gradient and electric energy densities to the total energy density.

The oscillation   behaviour   of   the   pressure  $\tilde{p}$  is  also rather
complicated.
Unlike the energy and Noether charge densities, the pressure $\tilde{p}$ may be
negative.
Indeed, we see that in Fig.~\ref{fig12b}, the  negative  minima  of $\tilde{p}$
are separated by positive peaks, the positions  of which approximately coincide
with those of the Noether  charge  density  $\tilde{j}_{N}^{0}$; these positive
peaks  are  due  to  the  kinetic  term  $\tilde{\Omega}^{2}\tilde{f}^{2}/2$ in
Eq.~(\ref{III:16b}).
The minima in $\tilde{p}$ have  a  double  well  structure  resulting  from the
interference of the last three terms in Eq.~(\ref{III:16b}).
The central maxima of these double wells approximately  coincide with the zeros
in $\tilde{j}_{N}^{0}$, whereas the two side minima of the double wells are due
to the contribution from  the negative gradient term $ -\tilde{f}^{\prime 2}/6$
in Eq.~(\ref{III:16b}).

The  solution   in   Fig.~\ref{fig12a}   corresponds   to  the  initial  values
$\tilde{f}_{0}=2.9$ and $\tilde{\Omega}_{0} = 0.6$; for a given $\tilde{e}$ and
$\tilde{h}$, it has the maximum possible  number of nodes ($n=9$).
Note that these  initial  values  satisfy  the condition of the quasi-thin-wall
regime~(\ref{IV:5a}).
We also studied radially excited $Q$-ball solutions with the same $\tilde{\Omega
}_{0}$ but a smaller number of nodes.
All these solutions  also  satisfy the quasi-thin-wall condition (\ref{IV:5a}),
and thus their initial values $\tilde{f}_{0}$ are  very  close  to  that of the
solution in Fig.~\ref{fig12a}.
It was found  numerically  that  starting  with $n = 4$, the energy and Noether
charge of these solutions satisfy the approximate relations
\begin{equation}
\tilde{E}^{1/3}\approx a_{\tilde{E}} + b_{\tilde{E}} n,\quad
\tilde{Q}_{N}^{1/3} \approx a_{\tilde{Q}_{N}}
+ b_{\tilde{Q}_{N}} n,                                             \label{IV:9}
\end{equation}
where the  coefficients  $a_{\tilde{E}}$, $b_{\tilde{E}}$, $a_{\tilde{Q}_{N}}$,
and $b_{\tilde{Q}_{N}}$ depend on the position of the  initial  point $(\tilde{
f}_{0}, \tilde{\Omega}_{0})$ on the quasi-thin-wall regime curve (\ref{IV:5a}).
Note  that  the  relations (\ref{IV:9})  are  similar  to  those   obtained  in
Ref.~\cite{mai_2012} for radially excited nongauged $Q$-balls.
In the latter case,  however,  the  number  of  radially  excited  $Q$-balls is
infinite, whereas  in  the  gauged  case,  it  is  finite  and increases with a
decrease in the gauge coupling constant $\tilde{e}$.

\section{\label{sec:V} Conclusion}

In the  present  paper,   radially  excited  $U(1)$  gauged $Q$-balls have been
investigated both analytically and numerically.
In particular, the  domain  of  existence  of   $U(1)$  gauged $Q$-balls in the
parameter space has been found by numerical methods.
It has also been found  that  the  presence  of  an   Abelian gauge field leads
to substantial changes in the properties  of  radially excited gauged $Q$-balls
compared to nongauged ones.
Firstly, there   exists  only  a  finite  number  of  radially  excited  gauged
$Q$-balls  at  given  values  of  the   model's   parameters,  whereas  in  the
nongauged case, the number of  radially excited $Q$-balls is infinite.
For the $n$-th   radially   excited  gauged  $Q$-ball  solution,  there exist a
maximum  possible Noether  charge  and  energy,  both of which decrease with an
increase in $n$.
For a given $n$,  there  also  exists  a  maximum  allowable value of the gauge
coupling constant; there  is  no $n$-th radially excited gauged $Q$-ball if the
gauge coupling constant exceeds this limiting value.
In the same way as the maximum possible Noether charge and energy,  the maximum
allowable gauge coupling constant decreases with an increase in $n$.
Note that this behaviour of  the  radially excited  gauged $Q$-balls is similar
to that of the unexcited one \cite{klee, gulamov_2015}.

Another  characteristic   feature  of   gauged  $Q$-balls (both  unexcited  and
radially  excited)  is   the  existence  of   turning   points  in  the  curves
describing  the  dependences  of  the  energy,  the  Noether charge and several
other quantities  on the parameter $\Omega_{\infty}$.
All of these turning points are  due  to  the  inflection  point  on  the curve
$E(Q_{N})$ describing  the  dependence  of  the  energy  of the $Q$-ball on the
Noether charge.
This inflection  point, in   turn,  results  from  the long-range nature of the
electrostatic  Coulomb repulsion.

Unlike nongauged $Q$-balls,   gauged ones (both unexcited and radially excited)
may have a negative pressure in the central domain.
This behaviour holds for the gauged $Q$-balls with a sufficiently large Noether
charge (quasi-thin-wall regime).
However, the  central  pressure  of  the gauged  $Q$-ball becomes positive with
increases in  both  the  gauge  coupling  constant  and  the radial excitation.

At a  fixed  Noether  charge,  the  energy  of  the  gauged  $Q$-ball increases
with an  increase in  the  radial  excitation  (i.e., with an increase in $n$).
It follows that the $i$-th  excited  gauged  $Q$-ball  can transition into less
excited $Q$-balls with $n = 1, \ldots, i-1$  or into an unexcited $Q$-ball with
$n = 0$; the   energy released is  carried  away by  electromagnetic and scalar
waves.

In the present paper,  we  have  considered  radially  excited gauged $Q$-balls
in  three-dimensional  space,   whereas    three-dimensional  radially  excited
nongauged $Q$-balls were  studied  in Refs.~\cite{volkov_2002, mai_2012}.
It is obvious that  radially  excited  nongauged  $Q$-balls  also  exist in two
spatial dimensions and do not exist in one dimension.
The existence of radially excited nongauged $Q$-balls  is  due to the existence
of the ``friction'' term $(d-1)f'(r)/r$   in  the differential equation for the
ansatz function $f(r)$.
This term is non-zero in the two-dimensional ($d= 2$) case, and vanishes in the
one-dimensional  ($d  =  1$)  case,  making  the  existence  of one-dimensional
radially excited $Q$-balls impossible.

It is known  that  one-  and  two-dimensional electrically charged objects must
have infinite electrostatic  energy,  meaning  that  the  existence  of one and
two-dimensional  electrically  charged  solitons  in  Maxwell  gauge  models is
impossible.
Note, however,  that  two-dimensional  electrically  charged solitons can exist
in gauge  models  for  which  the  Lagrangians  include  the  Chern-Simons term
\cite{paul, khare_rao_227, khare_255, hong, jw1, jw2, bazeia_1991, ghosh,
 deshaies_2006}.
In Maxwell gauge models, however, there  are  one-  and two-dimensional soliton
systems with zero  total  electric   charge   but  a  non-zero   electric field
\cite{loginov_plb_777, loginov_PRD_2019, loginov_EPJC_2019}.
These soliton systems are unexcited; however,  there are no reasons prohibiting
radial (in  two  spatial  dimensions)  or  linear (in  one  spatial  dimension)
excitations for them.

\begin{acknowledgments}

This work was supported by the Russian Science Foundation, grant No 19-11-00005.

\end{acknowledgments}

\appendix

\section{Time independence of gauged $Q$-ball fields in the unitary gauge}

To establish  the  time  dependence  of  the  gauged  $Q$-ball, the Hamiltonian
formalism  can  be  used  together   with   the   Lagrange  multipliers  method
\cite{fried1, fried2}.
First of all, we must fix the gauge.
We shall use the unitary gauge, in  which  the  imaginary  part  of the complex
scalar field vanishes:
\begin{equation}
\text{Re}\left( \phi \right) = \frac{f}{\sqrt{2}},\quad
\text{Im}\left( \phi \right) = 0,                                   \label{A:1}
\end{equation}
and the Lagrangian density (\ref{II:1}) takes the form
\begin{equation}
\mathcal{L}=-\frac{1}{4}F_{\mu \nu }F^{\mu \nu } + \frac{1}{2}
\partial_{\mu}f \partial^{\mu }f +
\frac{e^{2}}{2}A_{\mu}A^{\mu}f^{2} - V\left( f \right).             \label{A:2}
\end{equation}
Using  Eq.~(\ref{A:2}), we  obtain  the  generalized  momenta  corresponding to
the fields  $f$, $A_{x} = A^{1}$, $A_{y} = A^{2}$, and $A_{z} = A^{3}$:
\begin{subequations} \label{A:3}
\begin{eqnarray}
\Pi _{f} &=&\frac{\partial \mathcal{L}}{\partial \left( \partial
_{t}\phi _{1}\right) } = \partial_{t}f,                            \label{A:3a}
\\
\Pi _{A_{x}} &=&\frac{\partial \mathcal{L}}{\partial \left( \partial
_{t}A_{x}\right) }=F_{10}=-E_{x},                                  \label{A:3b}
\\
\Pi _{A_{y}} &=&\frac{\partial \mathcal{L}}{\partial \left( \partial
_{t}A_{y}\right) }=F_{20}=-E_{y},                                  \label{A:3c}
\\
\Pi _{A_{z}} &=&\frac{\partial \mathcal{L}}{\partial \left( \partial
_{t}A_{z}\right) }=F_{30}=-E_{z}                                   \label{A:3d}
\end{eqnarray}
\end{subequations}
and the Hamiltonian density
\begin{eqnarray}
\mathcal{H} &=&\partial_{t} f \Pi_{f} + \partial_{t}A_{x}\Pi
_{A_{x}}+\partial_{t}A_{y}\Pi _{A_{y}}+\partial _{t}A_{z}\Pi _{A_{z}}-
\mathcal{L}                                                      \nonumber
\\
&=&\frac{1}{2}\Pi _{A_{x}}^{2}+\frac{1}{2}\Pi _{A_{y}}^{2}+\frac{1
}{2}\Pi _{A_{z}}^{2}+\frac{1}{2}\Pi _{f}^{2}+\frac{1}{4}F_{ij}F_{ij}
                                                                 \nonumber
\\
&&+\frac{1}{2}\left( \partial _{x} f \right) ^{2}+\frac{1}{2}\left(
\partial_{y} f \right)^{2} + \frac{1}{2}\left( \partial_{z} f \right)^{2}
                                                                 \nonumber
\\
&&-\frac{e^{2}}{2} f^{2}\left(
A_{0}^{2}-A_{x}^{2}-A_{y}^{2}-A_{z}^{2}\right)                      \label{A:4}
\\
&&-\Pi _{A_{x}}\partial _{x}A_{0}-\Pi _{A_{y}}\partial _{y}A_{0}-\Pi
_{A_{z}}\partial _{z}A_{0}+V\left(f\right),              \nonumber
\end{eqnarray}
where we express the time derivatives of the fields in terms of the generalized
momenta.

Since the Hamiltonian  density  (\ref{A:4})  does  not depend on the derivative
$\partial_{t} A_{0}$, we have the primary constraint $\Pi _{A_{0}} = 0$.
The primary constraint must hold  at  any instant of time, and thus the Poisson
bracket of the generalized momentum $\Pi_{A_{0}}$ with the Hamiltonian $H =\int
\mathcal{H}d^{3}x$ must vanish.
This condition leads us to the secondary constraint
\begin{equation}
\frac{\delta H}{\delta A_{0}}=\partial_{x}\Pi_{A_{x}}+\partial_{y}\Pi_{A_{y}}+
\partial_{z}\Pi_{A_{z}}-e^{2}f^{2}A_{0} = 0.                        \label{A:5}
\end{equation}
Taking into account  the  expression  for  the  electric  charge density in the
unitary gauge: $\rho  =  j_{0} = -e^{2} f^{2}A_{0}$, and the definitions of the
generalized momenta in Eqs.~(\ref{A:3b}) -- (\ref{A:3d}), we see that constraint
(\ref{A:5}) is Gauss's law
\begin{equation}
\partial_{i}E_{i} = \rho.                                           \label{A:6}
\end{equation}
Gauss's law makes  it  possible  to  represent  the  Noether  charge of a field
configuration of  the  model  as  the  surface  integral  of  the  flux  of the
generalized momenta:
\begin{eqnarray}
Q_{N} &=&e^{-1}\int \rho d^{3}x
 = -e^{-1}\int \partial _{i}\Pi _{A^{i}}d^{3}x                   \nonumber
\\
&=&-e^{-1}\oint\nolimits_{S_{\infty }}\Pi_{A^{i}}dS_{i}.            \label{A:7}
\end{eqnarray}

The energy density $\mathcal{E}=T_{00}$ can also be  expressed  in terms of the
generalized momenta and the corresponding fields
\begin{eqnarray}
\mathcal{E} &=&\frac{1}{2}\Pi _{A_{x}}^{2}+\frac{1}{2}\Pi _{A_{y}}^{2}+\frac{
1}{2}\Pi_{A_{z}}^{2} + \frac{1}{2}\Pi_{f}^{2}+\frac{1}{4}F_{ij}F_{ij}
\nonumber
\\
&&+\frac{1}{2}\left( \partial _{x}f\right)^{2} + \frac{1}{2}\left(
\partial_{y}f\right)^{2} + \frac{1}{2}\left(\partial_{z}f \right)^{2}
\nonumber
\\
&&+\frac{e^{2}}{2} f^{2}\left(A_{0}^{2}+A_{x}^{2}+A_{y}^{2}+A_{z}^{2}\right)
+ V\left(f\right).                                                  \label{A:8}
\end{eqnarray}
We see  that  the  energy  density  (\ref{A:8})  does  not  coincide  with  the
Hamiltonian density (\ref{A:4}):
\begin{eqnarray}
\mathcal{E}-\mathcal{H} & = &\Pi _{A_{x}}\partial _{x}A_{0}+\Pi
_{A_{y}}\partial _{y}A_{0}                                       \nonumber
\\
& & + \Pi_{A_{z}}\partial_{z}A_{0} + e^{2}A_{0}^{2}f^{2}.           \label{A:9}
\end{eqnarray}
However, after integrating  by  parts,  the  right-hand side of Eq.~(\ref{A:9})
vanishes for field configurations  which  have   finite  energy (so $\underset{
r\rightarrow 0}{\lim }\Pi _{A^{i}} = 0$) and  satisfy  Gauss’s law (\ref{A:5}),
and hence for these field configurations
\begin{equation}
E = \int \mathcal{E}d^{3}x = H = \int \mathcal{H}d^{3}x.           \label{A:10}
\end{equation}

It can be shown that in  the  unitary  gauge, the field equations (\ref{II:6a})
and (\ref{II:6b}) can be written in the Hamiltonian form:
\begin{equation}
\partial_{t}\Pi_{f}=
-\frac{\delta H}{\delta f}=
-\frac{\delta E}{\delta f},\;
\partial_{t}\Pi_{A^{i}}=
-\frac{\delta H}{\delta A^{i}}=
-\frac{\delta E}{\delta A^{i}},                                    \label{A:11}
\end{equation}
where Eq.~(\ref{A:10}) is used.
When calculating  the  variational  derivatives  in  Eq.~(\ref{A:11}), the time
component $A_{0}$  of  the electromagnetic potential must be expressed in terms
of the canonical variables $f$ and $\Pi_{A^{i}}$ using Gauss's law (\ref{A:5}):
\begin{equation}
A_{0}=e^{-2}f^{-2}\left( \partial _{x}\Pi _{A_{x}}+\partial_{y}\Pi
_{A_{y}}+\partial _{z}\Pi _{A_{z}}\right).                         \label{A:12}
\end{equation}
The remaining Hamilton equations:
\begin{equation}
\partial_{t}f = \frac{\delta E}{\delta \Pi_{f}} =
\frac{\delta H}{\delta \Pi_{f}},\;
\partial_{t}A^{i} = \frac{\delta E}{\delta \Pi_{A_{i}}} =
\frac{\delta H}{\delta \Pi _{A^{i}}}                               \label{A:13}
\end{equation}
are  simply   the   definitions  of  the  generalized  momenta  (\ref{A:3a}) --
(\ref{A:3d}).
Next,  from   Eq.~(\ref{III:2}),   it   follows  that  in  the  $Q$-ball  field
configuration,  the  variation in the energy $\delta E = \lambda \delta Q_{N}$,
where $\lambda$ is the Lagrange multiplier.
In deriving the field equations (\ref{A:11}) and (\ref{A:13}), we must consider
only those variations of fields and generalized momenta  that vanish at spatial
infinity.
In particular, $\left. \delta \Pi _{A^{i}}\right\vert_{S_{\infty }}=0$, so from
Eq.~(\ref{A:7}) it follows that $\delta Q_{N} = 0$ for these variations.
Since $\delta E =\lambda \delta Q_{N}$ in the $Q$-ball field configuration, the
variation $\delta E$ also vanishes.
It then follows from Eqs.~(\ref{A:11}) and (\ref{A:13}) that the fields $f$ and
$A^{i}$  and the corresponding generalized momenta $\Pi_{f}$ and  $\Pi_{A^{i}}$
do not depend on time in the unitary gauge.
The time component $A_{0}$ of the electromagnetic potential also does not depend
on time in the unitary gauge, as this follows from Eq.~(\ref{A:12}).
Note, however, that  Eq.~(\ref{III:2}) also holds  for variations in the fields
and generalized  momenta  that lead  to non-zero $\delta E$ and $\delta Q_{N}$.
In particular, it  holds  for  variations  connecting two infinitesimally close
$Q$-ball  field  configurations,  giving  rise  to  the  differential  relation
(\ref{III:3}).

Now let us investigate how the Lagrange multiplier  $\lambda$ is related to the
parameters of the gauged $Q$-ball.
In order to do this,  we  shall follow the method of  Ref.~\cite{gulamov_2014}.
Firstly,  note  that  in  the  unitary  gauge,  the  electromagnetic  potential
$A_{0}$ of the gauged $Q$-ball must tend  to some constant value, otherwise the
energy of the gauged $Q$-ball would be infinite.
Let us  denote this limiting value of $A_{0}\left( r \right)$  as $-e^{-1}\bar{
\lambda}$; then, the combination $\Omega \left(r\right)=-e A_{0}\left(r\right)$
can be represented as
\begin{equation}
\Omega \left(r\right)=\bar{\Omega}\left(r\right) + \bar{\lambda},  \label{A:14}
\end{equation}
where $\underset{r \rightarrow \infty }{\lim}\bar{\Omega }\left( r \right) =0$.
Using  Eqs.~(\ref{III:14}),  (\ref{III:15a}),  and  (\ref{A:14}),   we   obtain
expressions for derivatives of the Noether charge $Q_{N}$ and the energy $E$ of
the gauged $Q$-ball with respect to $\bar{\lambda}$:             
\begin{equation}
\frac{dQ_{N}}{d\bar{\lambda }}=4\pi \int \limits_{0}^{\infty}
\left(f^{2}\left(1 + \frac{d\bar{\Omega }}{d\bar{\lambda }}\right)
+ 2 f \frac{d f}{d\bar{\lambda }}\left( \bar{\Omega }+\bar{\lambda
}\right) \right) r^{2}dr,                                          \label{A:15}
\end{equation}
\begin{eqnarray}
\frac{dE}{d\bar{\lambda}} &=&4 \pi \int \limits_{0}^{\infty}
\left( e^{-2}\bar{\Omega }
^{\prime }\left( \frac{d\bar{\Omega }}{d\bar{\lambda }}\right)
^{\prime }+f^{\prime }\left( \frac{d f}{d\bar{\lambda}}
\right) ^{\prime }\right.                                        \nonumber
\\
&&+\left. f^{2}\left( \bar{\Omega }+\bar{\lambda }\right)
\left( 1+\frac{d\bar{\Omega }}{d\bar{\lambda }}\right) \right.   \nonumber
\\
&&+\left. f \frac{d f}{d\bar{\lambda }}\left( \bar{\Omega }+\bar{\lambda }
\right)^{2}+\frac{dV}{df}\frac{df}{d\bar{\lambda}}\right)r^{2}dr,  \label{A:16}
\end{eqnarray}
where the prime means differentiation with respect  to the radial variable $r$.
Next, we integrate the term $f^{\prime}\left(df /d\bar{\lambda}\right)^{\prime}
r^{2}$ in Eq.~(\ref{A:16}) by parts and use Eqs.~(\ref{III:5a}), (\ref{III:6}),
and (\ref{A:15}) to recast Eq.~(\ref{A:16}) in the form
\begin{eqnarray}
\frac{dE}{d\bar{\lambda }}
&=&\bar{\lambda}\frac{dQ_{N}}{d\bar{\lambda}}+4\pi \int\limits_{0}^{\infty}
\left(e^{-2}\bar{\Omega }^{\prime }\left( \frac{d\bar{\Omega}}
{d \bar{\lambda }}\right)^{\prime }\right.                         \label{A:17}
\\
&&+\left. \bar{\Omega }\left(f^{2}\left(1 + \frac{d\bar{
\Omega }}{d\bar{\lambda }}\right) + 2 f \frac{d f}
{d\bar{\lambda }}\left( \bar{\Omega } + \bar{\lambda}\right)
\right) \right) r^{2}dr.                                         \nonumber
\end{eqnarray}
We can then  use  Eq.~(\ref{A:14})  and  differentiate  Eq.~(\ref{III:5b}) with
respect to $\bar{\lambda}$ to obtain the relation
\begin{eqnarray}
& & \left( \frac{d\bar{\Omega }}{d\bar{\lambda }}\right) ^{\prime
\prime }+\frac{2}{r}\left( \frac{d\bar{\Omega }}{d\bar{\lambda }}
\right) ^{\prime }-e^{2} f^{2}\left( 1+\frac{d\bar{\Omega }}
{d\bar{\lambda }}\right)                                         \nonumber
\\
& & - 2 e^{2} f \frac{d f}{d\bar{\lambda }}
\left( \bar{\Omega } + \bar{\lambda }\right) = 0.                  \label{A:18}
\end{eqnarray}
We now  integrate  the  term $e^{-2}\bar{\Omega}^{\prime} \left( d\bar{\Omega}/
d\bar{\lambda }\right)^{\prime }r^{2}$   in  Eq.~(\ref{A:17}) by  parts and use
Eq.~(\ref{A:18}) to show that in Eq.~(\ref{A:17}),  the  integral term vanishes
for the $Q$-ball field configuration.
Thus  $dE/d\bar{\lambda}   =  \bar{ \lambda}dQ_{N}/d\bar{\lambda}$,   and  as a
consequence we obtain the relation
\begin{equation}
\frac{dE}{dQ_{N}} = \bar{\lambda}.                                 \label{A:19}
\end{equation}
From Eqs.~(\ref{III:3}) and (\ref{A:19}) it follows that $\lambda= \bar{\lambda
}$, and we conclude that in the unitary gauge, the Lagrange multiplier $\lambda
= - e A_{0}^{\infty}$,   where  $A_{0}^{\infty}$  is  $\underset{ r \rightarrow
\infty}{\lim }A_{0}\left( r \right)$.

We have shown that the  gauge  $Q$-ball solution does not depend on time in the
unitary gauge.
Of course,  this  time  independence  is  not gauge invariant.
In particular, under  the  gauge  transformations  (\ref{II:3})  with the gauge
function $\Lambda\left( \mathbf{x}, t \right) = e^{-1} \omega t$,  the  unitary
$Q$-ball solution (\ref{III:1}) becomes the gauge transformed solution
\begin{subequations} \label{A:20}
\begin{eqnarray}
\phi ^{\omega }\left( r,t\right)  &=&f\left( r\right) \exp \left( -i\omega
t\right),                                                         \label{A:20a}
\\
A_{0}^{\omega }\left( r\right)  &=&A_{0}\left( r\right)
+ e^{-1}\omega,                                                   \label{A:20b}
\end{eqnarray}
\end{subequations}
which depends on $t$.
At the same time, the  combination  $\Omega^{\omega}\left(r\right) = \omega - e
A_{0}^{\omega} \left( r\right)$ does not depend on the gauge parameter $\omega$
and thus coincides with the unitary gauge  combination  $\Omega\left(r\right) =
-e A_{0}\left(r\right)$.
In particular,  it  follows from   Eq.~(\ref{A:20b})    that $\Omega_{\infty} =
\underset{r \rightarrow \infty}{\lim}\Omega^{\omega}\left(r\right)$ is equal to
the Lagrange multiplier $\lambda$:
\begin{eqnarray}
\Omega _{\infty } &=&\underset{r\rightarrow \infty }{\lim }
\left( \omega - e A_{0}^{\omega }\left( r \right) \right)    \nonumber
\\
&=&-e\underset{r\rightarrow \infty }{\lim }A_{0}\left( r\right)
=- e A_{0}^{\infty } = \lambda,                                    \label{A:21}
\end{eqnarray}
and thus the  differential  relation  (\ref{III:3})  can be written in the form
\begin{equation}
\frac{dE}{dQ_{N}} = \Omega_{\infty},                               \label{A:22}
\end{equation}
which is valid for an arbitrary gauge parameter $\omega$.
Note that if $\underset{r \rightarrow \infty}{\lim}A_{0}^{\omega}\left(r\right)
= 0$ (this gauge is  often  used  to describe gauged $Q$-balls), then $\Omega_{
\infty } = \omega$  and  Eq.~(\ref{A:22})  takes the form $dE/dQ_{N} = \omega$.

\section{Existence of  a maximum possible electric charge for gauged $Q$-balls}

It is known that in most cases, the electric charge of a gauged $Q$-ball cannot
be arbitrarily large.
It was  pointed  out  in  Ref.~\cite{tamaki_2014}  that  this  situation arises
when the  second  derivative  of  the  self-interaction potential $d^{2}V\left(
\left\vert \phi \right\vert\right)/d\left\vert\phi\right\vert^{2}$ is finite at
$\left\vert \phi \right\vert=0$.
We will discuss this point in more detail.
Multiplying  Eq.~(\ref{III:5b})  by $r$ and integrating by parts, we obtain the
integral relation $\int \nolimits_{0}^{\infty}\Omega^{\prime }\left(r\right) dr
= e \int \nolimits_{0}^{\infty} r j_{0}\left( r \right) dr$, which implies that
\begin{equation}
\Delta \Omega =\Omega_{\infty} - \Omega_{0}
=e\int \limits_{0}^{\infty } r j_{0} \left( r\right) dr.            \label{B:1}
\end{equation}
On  the  other  hand,  Eqs.~(\ref{III:7a}), (\ref{III:11}), (\ref{III:12}), and
(\ref{III:13}) lead us to the conclusion
\begin{equation}
\Delta \Omega < \Omega_{\infty} \le m,                              \label{B:2}
\end{equation}
where we  assume  for  definiteness  that  the  parameter  $\Omega_{\infty}$ is
positive.
Eqs.~(\ref{B:1}) and (\ref{B:2}) result in the inequality
\begin{equation}
e \int \limits_{0}^{\infty } r j_{0}\left(r\right) dr < m,          \label{B:3}
\end{equation}
where for field models  with  regular  self-interaction potentials, the squared
mass  $m^{2} = 2^{-1}d^{2}V\left(\left\vert \phi \right\vert\right)/d\left\vert
\phi\right\vert^{2}$ at $\left\vert \phi \right\vert = 0$.

We now show that Eq.~(\ref{B:3}) cannot  be  satisfied  if  the gauged $Q$-ball
can possess an arbitrarily large electric charge  $Q=4\pi\int_{0}^{\infty}r^{2}
j_{0}\left(r\right) dr$.
Firstly,  it  should  be  noted  that the electric charge density $j_{0}\left(r
\right) = e \Omega \left( r\right) f\left( r \right)^{2}$ is a bounded function
of $r$.
Indeed, from Eq.~(\ref{III:7}) it follows that $\Omega\left(r\right) \in \left(
0, m \right]$ and thus is bounded.
At the same time, it can be shown  that  regardless  of  the  value of $r$, the
ansatz  function  $f\left( r \right)$   cannot   exceed   the   limiting  value
$\sqrt{2g/h}$; otherwise, the term $(\Omega^{2} - m^{2})f+gf^{3}/2-hf^{5}/4$ in
Eq.~(\ref{III:5a})  is  negative  for $\Omega \in \left(0, m \right]$  and thus
$f\left(r\right)$ increases indefinitely.
The  boundary   condition   $f\left( r \right) \underset{ r \rightarrow \infty}
{\longrightarrow} 0$  is  therefore  not  met, and the gauged $Q$-ball does not
exist.
We can conclude that  the  electric  charge  density  of the gauged $Q$-ball is
bounded on the interval $r \in \left[0, \infty \right)$.

Now we consider the possible variants  of  the  electric  charge density $j_{0}
\left(r\right)$  that lead  to an arbitrarily large electric  charge $Q = 4 \pi
\int\nolimits_{0}^{\infty}j_{0}\left(r\right) r^{2}dr$.
The  first  variant  corresponds  to  the  case  when  the  integral  $I = \int
\nolimits_{0}^{\infty}j_{0}\left(r\right) dr$ is infinite.
In this case, however, the integral $J=\int\nolimits_{0}^{\infty} r j_{0}\left(
r \right) dr$  is  also infinite and thus Eq.~(\ref{B:3})  cannot be satisfied.
Consequently,  this  variant  of  $j_{0}\left( r \right)$  cannot  be realized.

The second variant of $j_{0}\left( r \right)$  corresponds to the case when the
integral $I=\int\nolimits_{0}^{\infty}j_{0}\left(r\right) dr$ is finite but the
electric charge $Q = 4\pi \int\nolimits_{0}^{\infty}j_{0}\left(r\right)r^{2}dr$
is infinite.
Note in  this connection that Eqs.~(\ref{III:11}) -- (\ref{III:13}) lead to the
conclusion that the electric charge density $j_{0}\left(r\right)=e\Omega \left(
r\right) f\left( r \right)^{2}$ tends  to  zero exponentially as $r \rightarrow
\infty$.
It follows that the electric charge $Q$  cannot  have  an infinite contribution
from the spatial asymptotics of $j_{0}\left( r \right)$.
If we suppose that the maximum of $j_{0}\left( r\right)$ does not tend to zero,
then this means that the electric charge  density  is localized within a finite
vicinity of this maximum, and  the  radial  position  of  the maximum increases
indefinitely.
Using the mean value theorem, we find that  for  this  electric charge density,
$Q \sim R^{2}$ and  $J \sim R$, where $R$ is the radial position of the maximum
of $j_{0}\left(r\right)$.
We see that an indefinite  increase in  $Q$  leads to an indefinite increase in
$R$, and as a consequence to an indefinite  increase  in the integral $J = \int
\nolimits_{0}^{\infty} r j_{0}\left( r \right) dr$.
Due to this, Eq.~(\ref{B:3}) again  cannot  be  satisfied,  and this variant of
$j_{0}\left(r\right)$ is also unrealizable.

There is  one  more  possible  variant, which corresponds to the case where the
electric charge density $j_{0}\left( r \right)$  spreads over the semi-infinite
interval $r \in \left[0, \infty\right)$  in  such  a  way that the integral $I$
tends to zero while the electric charge $Q$ increases indefinitely.
In this case, the integral $J=\int \nolimits_{0}^{\infty} r j_{0}\left(r\right)
dr$ may be finite, and condition (\ref{B:3}) may be met.
However, since the integral $I=\int \nolimits_{0}^{\infty} j_{0} \left(r\right)
dr$ tends to zero, the maximum of $j_{0}\left(r\right)$ also tends to zero, and
consequently  so  does  the maximum of the ansatz function $f\left( r \right)$.
We now show  that  this  behaviour  of $f\left( r \right)$  cannot be realized.
Firstly, note  that  the  system  of  differential equations (\ref{III:5a}) and
(\ref{III:5b}) can be represented in the form
\begin{subequations} \label{B:4}
\begin{eqnarray}
f^{\prime \prime }+\frac{2}{r}f^{\prime } &=&-\frac{\partial }{\partial f}
U_{\mathrm{eff}}\left( f,\Omega \right),                           \label{B:4a}
\\
\Omega ^{\prime \prime }+\frac{2}{r}\Omega ^{\prime } &=&-\frac{\partial }
{\partial \Omega }U_{\mathrm{eff}}\left( f,\Omega \right) +(1+e^{2})\Omega
f^{2},                                                             \label{B:4b}
\end{eqnarray}
\end{subequations}
where the effective potential
\begin{equation}
U_{\text{eff}}\left( f,\Omega \right) =\frac{1}{2}
\left(\Omega^{2}-m^{2}\right) f^{2}+\frac{g}{8}f^{4}-\frac{h}{24}f^{6}.
                                                                    \label{B:5}
\end{equation}
The system (\ref{B:4}) describes  the two-dimensional motion of a particle with
unit mass in the  plane $\left(f,\Omega\right)$, where the radial  variable $r$
plays the role of time.
The particle moves  in  a  viscous  medium  under  the action of a conservative
force $\mathbf{F}_{\text{c}} = \left(-\partial_{f} U_{\mathrm{eff}},-\partial_{
\Omega } U_{\mathrm{eff}} \right)$,   a   nonconservative  force  $\mathbf{F}_{
\text{nc}} = \left(0, (1 + e^{2}) \Omega f^{2} \right)$, and  a  friction force
$\mathbf{F}_{\text{f}} = \left(-2 r^{-1} f^{\prime }, -2 r^{-1} \Omega^{\prime}
\right)$, which is also nonconservative.

The structure of the level lines  of  the  effective  potential $U_{\text{eff}}
(f,\Omega)$ leads to the conclusion that the particle  must start moving in the
close vicinity of  the  point $(0, m)$  in  order  to  have  the  infinitesimal
``coordinate" $f\left( r \right)$.
In this case,  the  initial ``coordinate"  $\Omega_{0}$ satisfies the condition
\begin{equation}
m^{2} - \Omega_{0}^{2} \equiv \mu ^{2} \ll m^{2},                   \label{B:6}
\end{equation}
from which it follows that
\begin{equation}
\Delta \Omega = \Omega_{\infty} - \Omega_{0} < m - \Omega_{0} \approx
\frac{\mu^{2}}{2m}.                                                 \label{B:7}
\end{equation}

We now estimate the effective  radial  size  $\Delta R$  of the electric charge
distribution for this case.
From Eq.~(\ref{B:4a}), it  follows  that  the  motion of the particle along the
``coordinate" $f$ is determined  by  the action of the force $- \partial_{f}U_{
\text{eff}}$.
The effective potential $U_{\text{eff}}$  is  negative  in  the interval $f \in
\left(0,\, 2 g^{-1/2} \mu \right)$ and has  a  local  minimum  at  $f_{\min } =
2^{1/2}g^{-1/2}\mu$, where condition (\ref{B:6}) is used.
In the neighborhood of the local minimum $f_{\min}$,  the  effective  potential
$U_{\text{eff}}$ takes the form
\begin{equation}
U_{\mathrm{eff}} \approx - \frac{\mu^{4}}{2g} + \mu^{2}\left(f -
\sqrt{\frac{2}{g}}\mu\right)^{2},                                   \label{B:8}
\end{equation}
and thus  in  the  neighborhood  of  $f_{\min}$,  the motion of the particle is
harmonic with period $T = \sqrt{2}\pi/\mu$.
We can use the period $T$ to estimate $\Delta R$.
In doing  so,  we  neglect  the  friction  force $- 2 r^{-1}f^{\prime}$ and the
non-harmonic nature  of  the  effective potential $U_{\text{eff}}$, and assumed
that the  ``coordinate"  $\Omega$  is  fixed  when  the  particle  moves  along
the ``coordinate" $f$.
All these factors, however, can only  result  in  an  increase in the effective
radial size $\Delta R$, and thus we have the estimation
\begin{equation}
\Delta R > a \mu^{-1},                                              \label{B:9}
\end{equation}
where $a$ is a finite positive constant.
To estimate the value of the ansatz function $f$ on the interval $\Delta R$, we
note that the minimum point $f_{\min}= 2^{1/2}g^{-1/2}\mu$ and the nearest zero
point $f = 2g^{-1/2}\mu$ of the effective  potential $U_{\text{eff}}$ are  both
of the order $\mu$.
Hence, the  particle  moving  in  the  force  field  of the effective potential
$U_{\text{eff}}$ will have  the  ``coordinate"  $f$  of the order $\mu$  on the
``time" interval  $\Delta R$,  and we obtain the estimation
\begin{equation}
f > b \mu,                                                         \label{B:10}
\end{equation}
where $b$ is a finite positive constant.
Eqs.~(\ref{III:7a}) and (\ref{B:10}) lead  to  an  estimation  for the electric
charge density $j_{0}$ on the interval $\Delta R$
\begin{equation}
j_{0} > e \Omega_{0} b^{2}\mu^{2}.                                 \label{B:11}
\end{equation}
We can now obtain an estimation for the integral $J$:
\begin{eqnarray}
J &=&\int\limits_{0}^{\infty }rj_{0}\left( r\right)
dr=\int\limits_{0}^{\Delta R}rj_{0}\left( r\right) dr+A          \nonumber
\\
&>& e \Omega_{0} b^{2} \mu ^{2} \int\limits_{0}^{\Delta R}rdr + A >
\frac{1}{2} e \Omega_{0} a^{2} b^{2},                              \label{B:12}
\end{eqnarray}
where we use the fact that the  integral $A=\int\nolimits_{\Delta R}^{\infty }r
j_{0}\left( r \right)dr$   is   finite,   since   from   Eqs.~(\ref{III:11}) --
(\ref{III:13})  it  follows   that   $j_{0}\left( r \right)$   tends   to  zero
exponentially as $r \rightarrow \infty$.
In the same way, it can be  shown  that  the  integral  $J$  is  less than some
positive finite value.
Hence, the integral $J$ is  finite as $\mu \rightarrow 0$.
On the other hand, Eqs.~(\ref{III:14}), (\ref{B:9}), and (\ref{B:11}) result in
a lower estimation for the electric charge $Q$:
\begin{equation}
Q > \frac{4\pi}{3} e \Omega_{0} a^{3} b^{2}\mu ^{-1}.              \label{B:14}
\end{equation}
Thus, the integral $J$ is  finite,  whereas  the  electric charge $Q$ increases
indefinitely as $\mu \rightarrow 0$.
At the same time, Eq.~(\ref{B:7}) tells us that $\Delta \Omega < \mu^{2}/\left(
2 m \right)$ and therefore vanishes as $\mu \rightarrow 0$.
It follows that the  condition  $\Delta\Omega = e J$ (Eq.~(\ref{B:1})) does not
hold as $\mu \rightarrow 0$,  and  thus  a  gauged $Q$-ball with an arbitrarily
large electric charge cannot exist in this case.

We have shown  that   gauged $Q$-balls with arbitrarily  large electric charges
cannot exist in model (\ref{II:1}).
The reason lies in the constraint $\Omega_{\infty}\le m$, from which it follows
that the difference $\Delta\Omega = \Omega_{\infty} - \Omega_{0} < m$.
Due to  this  last  inequality,  Eq.~(\ref{B:1})  can   not  be  satisfied  for
sufficiently large  electric  charges,  and thus the corresponding $Q$-balls do
not exist.

The regular self-interaction potential $V\left(\left\vert\phi\right\vert\right)$
must have a finite second order derivative $d^{2}V\left(\left \vert \phi \right
\vert\right)/d\left\vert \phi \right\vert^{2}$  at $\left\vert \phi \right\vert
=0$, and thus the mass $m$ of the complex scalar field $\phi$ is also finite in
this case, since $m^{2}= 2^{-1}d^{2}V\left(\left\vert \phi \right\vert \right)/
d\left\vert\phi \right\vert^{2}$ at $\left\vert \phi \right\vert = 0$.
The difference $\Delta\Omega = \Omega_{\infty} - \Omega_{0}$  therefore remains
bounded for  all $Q$-balls  with regular self-interaction potentials, and hence
such  $Q$-balls  cannot  possess arbitrarily large electric charges.
It follows that  there  is  a   maximum  allowable electric charge for a gauged
$Q$-ball with a regular self-interaction potential.
However,  as  shown  in Refs.~\cite{ardoz_2009, tamaki_2014}, gauged  $Q$-balls
also exist in models where the self-interaction  potentials  are not regular at
$\left\vert \phi \right\vert = 0$.
In particular, the second order derivative $d^{2}V\left(\left \vert \phi \right
\vert \right)/d\left\vert \phi \right\vert^{2}$  diverges  as  $\left\vert \phi
\right\vert \rightarrow  0$, and thus there is no upper bound on the difference
$\Delta\Omega = \Omega_{\infty} - \Omega_{0}$ in these models.
This results in  the  existence  of  gauged  $Q$-balls  with  arbitrarily large
electric charges \cite{ardoz_2009, tamaki_2014}.

Another consequence of Eq.~(\ref{B:3}) is  that  gauged  $Q$-balls cannot exist
if the gauge coupling constant $e$ exceeds some upper bound.
Indeed, Eq.~(\ref{B:3}) cannot  be satisfied  for sufficiently large $e$, since
as shown  above, the  integral $J = \int \nolimits_{0}^{\infty} r j_{0} \left(r
\right) dr$ cannot be arbitrarily small.
Radially excited gauged $Q$-balls also cannot exist if  the number of nodes $n$
of the ansatz function $f(r)$ increases indefinitely.
In this case, the integral $J=\int \nolimits_{0}^{\infty} r j_{0}\left(r\right)
dr$ also increases indefinitely, and  Eq.~(\ref{B:3}) cannot be satisfied.
Hence, there is only a finite number of  radially excited  gauged  $Q$-balls at
given values of the model’s parameters.

\section{Existence of an inflection point  on  the curve $E(Q_{N})$ in the
gauged case}

In the  present  paper, the curves in Figs.~\ref{fig3}, \ref{fig4}, \ref{fig5},
\ref{fig8}, \ref{fig10}, and \ref{fig11} have  turning points at certain values
of the parameter $\Omega_{\infty}$.
The  existence   of  these  turning   points  results  from  the  existence  of
inflection points on the corresponding curves $E(Q_{N})$.
Indeed, by  definition,  the  second  derivative  $d^{2}E/dQ_{N}^{2} = 0$ at an
inflection point; the  basic relation (\ref{III:3})  then  tells  us  that  the
derivatives $d\Omega_{\infty }/dQ_{N}$ and $d\Omega _{\infty }/dE$  also vanish
at the inflection point.
Hence,  the  derivatives $dQ_{N}/d\Omega_{\infty }$  and $dE/d\Omega_{\infty }$
are infinite at the inflection  point,  resulting  in  the turning points shown
in Figs.~\ref{fig5} and \ref{fig8}.
Next, we differentiate Eq.~(\ref{B:1}) with respect to  the parameter $\Omega_{
\infty}$.
Taking into account that at the turning point, the infinite derivative $dQ_{N}/
d\Omega_{\infty}$ results from the infinite derivative $\partial j_{0}/\partial
\Omega _{\infty}$, we conclude that the derivative $d\Omega_{0}/d\Omega_{\infty
}$ becomes infinite at  the  turning  point in accordance with Figs.~\ref{fig3}
and \ref{fig4}.
Finally, keeping in mind that the central pressure $p_{0}=\Omega_{0}^{2}f_{0}^{
2}/2 - V(f_{0})$, we conclude that the derivative  $dp_{0}/d\Omega_{\infty}$ is
also infinite at the turning point, which is consistent with  Figs.~\ref{fig10}
and \ref{fig11}.

Combining Eqs.~(\ref{III:3})  and  (\ref{III:26}),  we  obtain the differential
relation
\begin{equation}
\frac{dE}{dQ_{N}}-\frac{E}{Q_{N}}+\frac{2}{3Q_{N}}\left( E^{\left( G\right)
}-E^{\left( E\right) }\right) = 0.                                  \label{C:1}
\end{equation}
From this equation, it follows that  if  we  know how the electric and gradient
parts of the energy depend on the Noether charge $Q_{N}$,  we can determine the
dependence of  the  total  energy  $E$  of the $Q$-ball solution on the Noether
charge $Q_{N}$.
In   this   case,   Eq.~(\ref{C:1})  becomes a first order linear inhomogeneous
differential equation  that  can  be  solved  by  the  method  of  variation of
constants.
Of course, the exact forms of the  dependences  $E^{\left(E\right)}(Q_{N})$ and
$E^{\left( E \right)}(Q_{N})$  are  unknown; however, we  can  guess  the  main
features of these  dependences for sufficiently large $Q_{N}$ when the interior
part and the  edge  of  the  $Q$-ball  solution  are  clearly  distinguishable.
The electrostatic energy of a compact object possessing the Noether charge $Q_{
N}$ can be written as $E^{\left(E\right)}(Q_{N})=\alpha Q_{N}^{2}/(2R(Q_{N}))$,
where $R(Q_{N})$ is the  object's  effective  charge  radius, which  depends on
$Q_{N}$, and $\alpha = e^{2}/(4 \pi)$ is the fine-structure constant.
For a uniform distribution of the electric charge, the  effective charge radius
$R \propto Q_{N}^{1/3}$.
If the electric charge is concentrated in the spherical shell of radius $R$ and
thickness $\Delta$,  then  the effective charge radius $R \propto Q_{N}^{1/2}$.
For a gauged $Q$-ball,  we  have an intermediate situation and therefore expect
that $R \approx \varrho Q_{N}^{\gamma}$, where $\varrho$ is a positive constant
and the exponent $\gamma \in (1/3, 1/2)$.
Thus,  the  electrostatic  energy of the $Q$-ball solution is written as
\begin{equation}
E^{\left( E\right)}(Q_{N}) \approx \frac{\alpha}{2\varrho}
Q_{N}^{2-\gamma} \equiv aQ_{N}^{2 - \gamma}.                        \label{C:2}
\end{equation}
Next, we suppose that for sufficiently large $Q_{N}$,  the main contribution to
the gradient energy $E^{\left(G\right)}$  comes  from  the  edge  region of the
$Q$-ball.
In this case,  the  gradient  energy  of  the  $Q$-ball solution takes the form
\begin{equation}
E^{\left( G\right) }(Q_{N}) \approx 4 \pi R^{2} T =
4\pi \varrho ^{2}TQ_{N}^{2\gamma} \equiv b Q_{N}^{2\gamma},         \label{C:3}
\end{equation}
where $T$ is the surface tension.

Substituting Eqs.~(\ref{C:2}) and (\ref{C:3})  into  Eq.~(\ref{C:1}), we obtain
a first-order linear inhomogeneous differential equation that can be integrated
by the method of variation of constants.
The solution to this differential equation can be written as
\begin{equation}
E = c Q_{N}+\frac{2}{3}\frac{a Q_{N}^{2 - \gamma}}{(1 - \gamma)} +
\frac{2}{3}\frac{b Q_{N}^{2\gamma}}{(1 - 2\gamma )},                \label{C:4}
\end{equation}
where $c$ is the integration constant.
Note that from Eq.~(\ref{C:4}), it follows that  the positive constant $\gamma$
should be less than $1/2$; otherwise, the last term in  Eq.~(\ref{C:4}) becomes
negative or  diverges,  which  is  unacceptable  from a physical point of view.
Under this condition,  all  the  exponents  in   Eq.~(\ref{C:4})  are positive.
Next, using Eq.~(\ref{C:4}), we obtain the  first and second derivatives of the
$Q$-ball's energy with respect to the Noether charge:
\begin{eqnarray}
\frac{dE}{dQ_{N}} &=&c+\frac{2}{3}\frac{a (2-\gamma )Q_{N}^{1-\gamma }}
{(1-\gamma )}+\frac{4}{3}\frac{b \gamma Q_{N}^{2\gamma -1}}{(1-2\gamma )},
                                                                    \label{C:5}
\\
\frac{d^{2}E}{dQ_{N}^{2}} &=&\frac{2}{3}a(2-\gamma )Q_{N}^{-\gamma }
-\frac{4}{3}b\gamma Q_{N}^{2\gamma - 2}.                            \label{C:6}
\end{eqnarray}
By equating the second derivative (\ref{C:6})  to  zero, we obtain the value of
the Noether charge at the inflection point:
\begin{equation}
Q_{N,\text{infl}}=\left( \frac{2b\gamma }{a(2-\gamma )}\right)^{
\frac{1}{2 - 3\gamma }}.                                            \label{C:7}
\end{equation}

Formally, the existence of the inflection point is related to the fact that the
two terms in Eq.~(\ref{C:6}) have opposite signs.
This is because the exponent $2 - \gamma$ in Eq.~(\ref{C:2}) is larger than the
exponent $2 \gamma$  in Eq.~(\ref{C:3}),  due  to the multiplier $Q_{N}^{2}$ in
the electrostatic energy $E^{(E)}$.
In turn, the multiplier $Q_{N}^{2}$ is a consequence  of  the long-range nature
of the electrostatic Coulomb repulsion.
Thus, it can be said that the  existence  of the inflection point is due to the
long-range Coulomb repulsion.
Indeed, from Eq.~(\ref{C:7}) it follows that $Q_{N,\text{infl}}\propto e^{-2/(2
- 3 \gamma )}$.
Hence, $Q_{N,\text{infl}}  \rightarrow  \infty$  as $e \rightarrow 0$, which is
equivalent to the absence of an inflection point at zero $e$  when  there is no
Coulomb repulsion.

As mentioned above, the derivative $dE/dQ_{N}$  cannot  exceed  the mass $m$ of
the scalar $\phi$-boson.
However, Eq.~(\ref{C:5}) tells us that  $dE/dQ_{N} \rightarrow \infty$ as $Q_{N}
\rightarrow 0$.
Hence, Eq.~(\ref{C:4})  becomes  inapplicable  for  sufficiently small $Q_{N}$.
In  particular, it does not reproduce the cusp at the minimum possible $Q_{N}$.
This is because the partition of the $Q$-ball into  an  interior and an edge is
unclear for  small  enough  $Q_{N}$,  and thus Eqs.~(\ref{C:2}) and (\ref{C:3})
become inapplicable in this case.

When $Q_{N}$ reaches the maximum  possible  value $Q_{N,\max }$, the derivative
$dE/dQ_{N} = m$.
Combining this result with  Eq.~(\ref{C:5}) allows us  to express the parameter
$c$ in terms of $Q_{N,\max}$ and the rest of the parameters:
\begin{equation}
c = m - \frac{2a}{3}\frac{2-\gamma }{1-\gamma}Q_{N,\max}^{1-\gamma}
-\frac{4}{3}\frac{b \gamma}
{\left(1 - 2\gamma \right)}Q_{N,\max}^{-1 + 2\gamma}.               \label{C:8}
\end{equation}

Finally, note that  Eqs.~(\ref{C:2})  and  (\ref{C:3}) can be  considered as an
approximation that is only valid for sufficiently large $Q_{N}$.
In  particular, a more accurate description can  be  achieved if the parameters
$\varrho$ and $T$ in Eqs.~(\ref{C:2}) and  (\ref{C:3})  are  some  functions of
$Q_{N}$.
Nevertheless, we believe  that  the simplified approach used here clearly shows
the reason for the  existence  of  the inflection point on the curve $E(Q_{N})$
in the gauged case.

\bibliography{article}

\end{document}